\documentclass{ws-ijmpa}
\usepackage[super,compress]{cite}
\usepackage{graphicx}
\begin{document}
\markboth{Gaia Lanfranchi}{Rare b-hadron decays as probe of new physics}

%
\catchline{}{}{}{}{}
%

\title{Rare b-hadron decays as probe of new physics}

\author{Gaia Lanfranchi\footnote{Laboratori Nazionali di Frascati dell'INFN, Frascati (Rome), Italy}.}

\address{ Laboratori Nazionali di Frascati - Istituto Nazionale di Fisica Nucleare \\
via Enrico Fermi 40, 00044, Frascati (Rome), Italy \\
Gaia.Lanfranchi@lnf.infn.it}

\maketitle

\begin{history}
\received{5 April 2018}
\revised{16 April 2018}
\end{history}

\begin{abstract}
  The unexpected absence of unambiguous signals of New Physics at the TeV scale at
  the Large Hadron Collider puts today flavour physics at the forefront.
  In particular rare decays of $b$-hadrons represent a unique probe to challenge the Standard Model
  paradigm and test models of
  New Physics at a scale much higher than that accessible by direct searches.
  This article reviews the status of the field.
  
\end{abstract}

\ccode{PACS numbers: 13.20.He, 12.15.Mn, 12.60.Cn, 12.60.Fr, 12.60.Jv, 14.80.Sv}


\section{Introduction}
  With the discovery at the Large Hadron Collider (LHC) of the Higgs boson~\cite{Higgs_atlas, Higgs_cms}
  the last missing particle for the experimental validation of the Standard Model (SM) has been found.
  An additional and very important LHC result is that a large new high-scale territory has been explored
  and no unambiguous signal of New Physics (NP) has been found.  

  These results indicate that there might be NP with a direct and sizeable
  coupling to SM particles only at very high masses, unaccessible by direct searches at present colliders,
  but in principle accessible by precision measurements.
  Another possibility is that new particles are below the electro-weak (EW) scale and couple very
  weakly with the SM world and so far escaped detection.

  \vskip 2mm
  Flavour changing neutral current (FCNC) $b$-hadron decays are forbidden at the tree level
  and can therefore only proceed via loop topologies.
  New physics models, instead, can introduce additional couplings to new heavy mediators at both
  tree and loop level and these couplings could modify the values of the branching fractions and/or angular observables
  with respect to the SM predictions.
  Hence FCNC decays of $b$-hadrons are a powerful probe of physics beyond the SM that can be at a scale
  much higher than that currently accessible by direct searches.

  \vskip 2mm
  In the recent years a wealth of experimental data on rare $b$-hadron decays has been accumulated by the
  LHCb, ATLAS and CMS experiments at the LHC, and the
  measurements of the branching fractions and angular observables are challenging the CKM picture with unprecendented sensitivity.

  \vskip 2mm
  Interesting hints of deviations from the SM predictions have emerged from the analysis of
  rare decays mediated by $b \to s \ell^+  \ell^-$ transitions. They are related to measurements of
  branching fractions~\cite{lhcb_BR_BKst, lhcb_BR_BKst_2, lhcb_BR_BPhi, lhcb_BR_BLambda},
  ratios of branching fractions~\cite{lhcb_RK, lhcb_RKstar}, and angular distributions~\cite{lhcb_P5, belle_P5}.
  These anomalies, together with the observed deviations from $\tau/\mu$ (and $\tau/e$) universality in
  $ b \to c \ell \overline{\nu}$ charged
  currents~\cite{BaBar_RD_RDst_0, BaBar_RD_RDst, Belle_RD_03233, Belle_RDst_07923, Belle_RD,lhcb_RDst, lhcb_RDst_2},
  have triggered several theoretical speculations about possible NP
  interpretations~\cite{Bhattacharya, Alonso, Greljo_01705, Calibbi, Bauer,Fajfer, Barbieri, Das, Boucenna_01349, Beciveric_08501, DiLuzio_08450, Hiller, Bhattacharya2, Buttazzo, Barbieri2,Bordone, Crivellin_09226, Beciveric2, Cai,Megias, Buttazzo2, Bordone2}.

  \vskip 2mm
  This paper reviews the state-of-the-art of the search for NP in FCNC $b-$hadron decays mostly at the LHC experiments.
  The review is organized as follows: in Section~2 a brief summary of the theoretical
  framework for rare $b-$hadron decays is presented;
  recent results on leptonic and semi-leptonic decays mediated by $b \to s \ell^+ \ell^-$
  transitions\footnote{The inclusion of charge-conjugate processes is implied throughout.}
  are presented in Section~3 and Section~4, respectively.
  Searches for new very-weakly coupled light particles possibly produced in $b-$meson decays
  are discussed in Section~5.
  Some conclusions are drawn in Section~6 along with an outlook for the results expected in the coming years.

  \vskip 2mm
  All the results presented in this document, except for the update of the measurement of the
  branching fractions of the $B^0_s \to \mu^+ \mu^-$ and $B^0 \to \mu^+ \mu^-$ decays,
  are based on datasets collected by the LHCb, ATLAS and CMS experiments, during Run 1 (2010-2012).
  The Run 1 datasets consist of $\sim$3 fb$^{-1}$, $\sim$20 fb$^{-1}$ and $\sim$20$^{-1}$ of integrated luminosity
  collected at LHCb, ATLAS and CMS experiments, respectively, with proton-proton ($pp$) collisions
  at a centre-of-mass energy of $\sqrt{s}=7$~TeV and $8$~TeV. 
  
  LHCb has recently updated the analysis of the $B^0_s \to \mu^+ \mu^-$ and $B^0 \to \mu^+ \mu^-$
  rare decays~\cite{lhcb_bsmm} by adding to the Run 1 dataset  $\sim 1.4$ fb$^{-1}$ of
  integrated luminosity collected in the current run (Run2, 2015-2018) at  $\sqrt{s}$ = 13 TeV.
  
  To date the LHCb and ATLAS/CMS experiments have collected $\sim 4$ fb$^{-1}$ and $\sim 100$ fb$^{-1}$
  at $\sqrt{s}$ = 13 TeV,
  respectively, which correspond to an increase in the $b-$hadron production yield by at least
  a factor of $\sim$ 2 (LHCb) and $\sim$ 8 (ATLAS and CMS each) with respect to that collected in Run 1,
  once the increase of the $b\overline{b}$ production cross section with the $pp$ collisions
  centre-of-mass energy is taken into account.

  \section{Theory framework}
  \label{sec:theory}
  If the SM is a low-energy effective theory of a more fundamental theory valid at higher energy scales,
  the effective Hamiltonian can be described by the sum of local operators $O_i$ with different Lorentz structure, multiplied by their
  Wilson coefficients $C_i$, all evaluated at a renormalization scale $\mu$:

  \begin{equation}
    H_{\rm eff} = - { 4 G_F \over \sqrt{2} } V_{tb} V^*_{tq} {e^2 \over 16 \pi^2} \sum_i {(C_i Q_i + C'_i Q'_i) + h.c.}
    \label{eqn:Heff}
  \end{equation}

  where $G_F$ is the Fermi constant and $V_{tb} V_{\rm tq}$ ($q=d,s,b$) are CKM matrix elements.
  
  Among the dimension-six operators contributing to these transitions, the operators most sensitive
  to new physics contributions are:

  \[
  \begin{array} {ll}
    Q_7^{q(')}  =  {m_b \over e} ( \overline{q} \sigma_{\mu\nu} P_{R(L)} b ) F^{\mu\nu} & 
    Q_8^{q(')} = {m_b \over e} ( \overline{q} \sigma_{\mu\nu} P_{R(L)} T^a b ) G^{a\mu\nu}  \\ \nonumber
    Q_7^{q(')}  =  {m_b \over e} ( \overline{q} \sigma_{\mu\nu} P_{R(L)} b ) F^{\mu\nu} & 
    Q_8^{q(')} = {m_b \over e} ( \overline{q} \sigma_{\mu\nu} P_{R(L)} T^a b ) G^{a\mu\nu} \\ \nonumber
    Q_9^{q(')} = (\overline{q} \gamma_{\mu} P_{L(R)} b) (\overline{l} \gamma^{\mu} l) &
    Q_{10}^{q(')} = (\overline{q} \gamma_{\mu} P_{L(R)} b) (\overline{l} \gamma^{\mu} \gamma_5 l)  \\ \nonumber
    Q_S^{q(')} = (\overline{q} P_{L(R)} b) (\overline{l} l)  & 
    Q_P^{q(')} = (\overline{q} P_{L(R)} b) (\overline{l}\gamma_5  l)  \\ 
    Q^q_{L(R)} = (\overline{q} \gamma_{\mu} P_{L(R)} b ) (\overline{\nu} \gamma^{\mu} P_L \nu).  & \\
    \label{array:operators} 
  \end{array}
  \]

  Here the primes indicate operators that have quark chirality opposite to the SM one,
  $P_{L,R}$ denote left and right-hand chirality projections and $F^{\mu\nu}$ and $G^{a\mu\nu}$ are the
  electromagnetic and chromomagnetic field strength tensors, respectively. The electromagnetic
  and chromomagnetic dipole operators $Q^{(')}_7$ and $Q^{(')}_8$ contribute to radiative
  and semileptonic decays. The semileptonic operators $Q^{(')}_{10,S,P}$ contribute to leptonic
  and semileptonic decays, the operators $Q^{(')}_9$ only to semileptonic decays, and the operators
  $Q_{L,R}$ to decays with neutrinos in the final state.

  NP contributions from heavy particles can either modify the Wilson coefficients  of SM operators
  and/or generate new operators not present in the SM.

  \section{Leptonic decays}
  \label{sec:leptonic}

  \noindent
  Leptonic decays
  $B^0_{(s)} \to \ell^+ \ell^-$  with $\ell  = e, \mu, \tau$ are among the most important
  indirect probes of NP at the LHC, as they
  are strongly suppressed in the SM, very sensitive to NP effects, and theoretically very clean.

  These decays are extremely rare in the SM because they are not only loop and CKM suppressed but also helicity
  suppressed as the two spin-1/2 leptons originate from a pseudo-scalar $B$ meson.
  The branching fractions of these decays can be written as:

  \begin{equation}
    \text{BR}(B^0_{(s)}\to\ell^+\ell^-)_\text{SM} =
    \tau_{B^0_{(s)}}
    \frac{G_F^2\alpha_\text{em}^2}{16\pi^2}
    f_{B^0_{(s)}}^ 2
    m_\ell^2 m_{B^0_{(s)}} \sqrt{1-\frac{4 m_\ell^2}{m_{B^0_{(s)}}^2}}
    |V_{tb}V_{tq}^*|^2 |C_{10}^\text{SM}|^2 \,,
    \label{eq:br-bll-sm}
  \end{equation}
  where $f_{B^0_{(s)}}$ is the $B^0_{(s)}$ meson decay constant,  $m_{\ell}$ and $m_{B^0_{(s)}}$ are the masses of
  the lepton and the $B$ meson, respectively, and $|V_{tb} V^{*}_{tq}|$ are the relevant CKM matrix elements.
  The SM branching fractions only depend on the Wilson coefficient $C_{10}$, whose contribution is
  suppressed (due to helicity suppression) by $m^2_{\ell}/ m^2_{B^0_{(s)}}$.
  The smallness of the electron and muon masses renders the branching fractions of the $B^0_{(s)} \to e^+ e^-$ and
  $B^0_{(s)} \to  \mu^+ \mu^-$  decay modes more strongly helicity suppressed than
  that of the $B^0_{(s)} \to \tau^+ \tau^-$ modes.

  \vskip 2mm
  Even if rare, these branching fractions are extremely well predicted: in fact, accounting for NLO
  electroweak corrections~\cite{bsmm_EW} and NNLO QCD corrections~\cite{bsmm_QCD1, bsmm_QCD2}
  to $C_{10}$, the total uncertainty is  $\sim 10\%$,  shared almost evenly between the knowledge
  of the $B^0$ and $B^0_s$ meson decay constants from lattice QCD~\cite{bsmm_fb1, bsmm_fb2, bsmm_fb3},
  and the CKM matrix elements. The current theory predictions:
  
  \begin{align}
    \overline{\text{BR}}(B^0_s\to e^+ e^-)_{\rm SM} &= (8.24 \pm 0.36)\times 10^{-14} \nonumber \\
    \text{BR}(B^0\to e^+ e^-)_{\rm SM} &= (2.63 \pm 0.32)\times 10^{-15}
    ,\label{eq:bsee_SM} 
  \end{align}

  \begin{align}
    \overline{\text{BR}}(B^0_s\to\mu^+\mu^-)_{\rm SM} &= (3.52 \pm 0.15)\times 10^{-9} \nonumber \\
    \text{BR}(B^0\to\mu^+\mu^-)_{\rm SM} &= (1.12 \pm 0.12)\times 10^{-10}
    ,\label{eq:bsmumu_SM} 
  \end{align}

  \begin{align}  
    \overline{\text{BR}}(B^0_s\to\tau^+\tau^-)_{\rm SM} &= (7.46 \pm 0.30)\times 10^{-7} \nonumber \\
    \text{BR}(B^0\to\tau^+\tau^-)_{\rm SM} &= (2.35 \pm 0.24)\times 10^{-8}.
    \label{eq:bstautau_SM} 
  \end{align}

  are taken from Ref.~\refcite{rarebdecays} and the notation $\overline{\text{BR}}$
  denotes the time-integrated branching fractions measured by the experiments,
  which for the $B^0_s$ decays, is different from the prompt one (Eq.~\ref{eq:br-bll-sm})
  due to the not negligible lifetime difference between the $B^0_s$ heavy and the light mass eigenstates,
  $\Delta \Gamma = (0.082 \pm 0.007)$ ps$^{-1}$~\cite{lhcb_bsmm}.
  
  The relation between the time-integrated branching fraction and the prompt one is given by:
  
  \begin{equation}
    \overline{\text{BR}}(B^0_s\to\ell^+\ell^-)
    =
    \left[
      \frac{1+\mathcal A^{\ell^+ \ell^-}_{\Delta\Gamma} \, y_s}{1-y_s^2}
      \right]
    \text{BR}(B^0_s\to\ell^+\ell^-) \,
    \label{eq:bsmm_DG}
  \end{equation}

  where $y_s = \Delta \Gamma_s/2\Gamma_s = 0.062 \pm 0.006$~\cite{hflag, pdg} and $\mathcal A_{\Delta \Gamma}$ is the
  mass eigenstates rate asymmetry, defined as
  $\mathcal A_{\Delta \Gamma} = -2 \mathcal R (\lambda)/(1+|\lambda|^2)$, with
  $\lambda = (q/p) (A(\overline{B}_s \to \mu^+ \mu^-)/A(B^0_s \to \mu^+ \mu^-))$. The complex coefficients $p$ and $q$
  define the mass eigenstates of the $B^0_s - \overline{B}^0_s$ system in terms of the flavour eigenstates,
  and $A(B^0_s \to \mu^+ \mu^-) (A(\overline{B}^0_s \to \mu^+ \mu^-))$ is the $B^0_s (\overline{B}_s)$ decay amplitude.
  $\mathcal A_{\Delta \Gamma} =1 $ in the SM.

  \vskip 2mm
  NP contributions arising from scalar or pseudo-scalar operators could lift the helicity
  suppression possibly enhancing the value of the branching fractions.
  Hence these modes are particularly sensitive to models with an extended Higgs sector,
  as in a minimal supersymmetric extension of the SM with two Higgs doublets~\cite{bsmm_2HDM_1, bsmm_2HDM_2}
  or in supersymmetric models with non-universal Higgs masses~\cite{bsmm_BSM}.
  Even in the absence of scalar or pseudoscalar operators, the rare decays $B^0_{s} \to \ell^+ \ell^-$ provide
  strong constraints on models predicting NP contributions in $C_{10}$ or a non-zero $C'_{10}$, that
  contribute also to semi-leptonic $b \to q \ell^+ \ell^-$ transitions. These operators can receive contributions from an effective
  flavour-changing Z couplings and/or by the exchange of
  a new heavy neutral vector boson $Z'$~\cite{Buras_1896} or various types of scalar or vector
  leptoquarks~\cite{dosner_04993}.

  \vskip 2mm
  In the SM, only the heavy state decays to $\mu^+ \mu^-$ but this condition does not necessarily hold in
  NP scenarios~\cite{bsmm_debruyn}.
  The contributions from the two states can be disentangled by measuring the $B^0_s\to \mu^+ \mu^-$
  effective lifetime which, in the search for physics beyond the SM,
  is a complementary probe to the branching fraction measurement.

  \vskip 2mm  
  The huge $b$ quark production at the LHC allowed the LHCb experiment to reach the first
  evidence~\cite{bsmm_evidence} for the decay $B^0_s \to  \mu^+ \mu^-$ in 2012 and then the LHCb~\cite{bsmm_lhcb_obs}
  and CMS~\cite{bsmm_cms_obs} collaborations to observe this mode separately in 2013, based on
  the full data sets collected in Run 1.
  A subsequent joint analysis~\cite{bsmm_combined} performed by LHCb and CMS boosted the observation
  of the $B^0_s \to \mu^+ \mu^-$ mode with a statistical significance of 6.2 standard deviations ($\sigma$)
  and showed an intriguing $3.0 \,\sigma$ evidence for the $BR(B^0 \to \mu^+ \mu^-)$ mode.
  The two central values were compatible with the SM predictions at $1.2 \,\sigma$ and $2.2 \,\sigma$ level
  for the $B^0_s \to \mu^+ \mu^-$ and $B^0 \to \mu^+ \mu^-$ modes, respectively.
  The ATLAS collaboration has also searched for the $B_s \to \mu^+ \mu^-$ mode~\cite{bsmm_atlas}
  obtaining a $BR(B^0_s \to \mu^+ \mu^-)= (0.9^{+1.1}_{-0.8}) \times 10^{-9}$.

\vskip 2mm
  In 2017 the LHCb collaboration presented a new result based on the full Run 1 data
  set  and 1.4 fb$^{-1}$ of $pp$ collisions
  collected at a $\sqrt{s}$ = 13~TeV in Run 2~\cite{bsmm_lhcb_obs}. 
  The $B^0_s \to \mu^+ \mu^-$ decay mode has been observed with a statistical significance of $7.8 \,\sigma$
  and represents the first single-experiment observation of the decay, with the branching fraction
  value of $BR(B^0_s \to \mu^+ \mu^-) = (3.0 \pm 0.6 ({\rm stat}) ^{+0.3}_{-0.2} ({\rm syst})) \times 10^{-9}$.  
  The enhancement of the $B^0 \to \mu^+ \mu^-$ branching fraction with respect to the SM
  predictions~\cite{bsmm_combined} has not been confirmed and an
  upper limit has been set, $BR(B^0 \to \mu^+ \mu^-) <4.2 \times 10^{-10}$ at 95\% confidence level (CL).

  \vskip 2mm
  LHCb also performed the first measurement of the effective lifetime~\cite{lhcb_bsmm},
  via the fit of the decay time distribution of the untagged sample:
  \begin{equation}
    \tau(B^0_s \to \mu^+ \mu^-) = 2.04 \pm 0.44 \pm 0.05 \;{\rm ps}
    \label{tau_bsmm}
  \end{equation}
  albeit the measurement has still
  large uncertainties that do not allow to contrain significantly $\mathcal A_{\Delta \Gamma}$.

\begin{figure}[htb]
     \centerline{\includegraphics[width=0.75\linewidth]{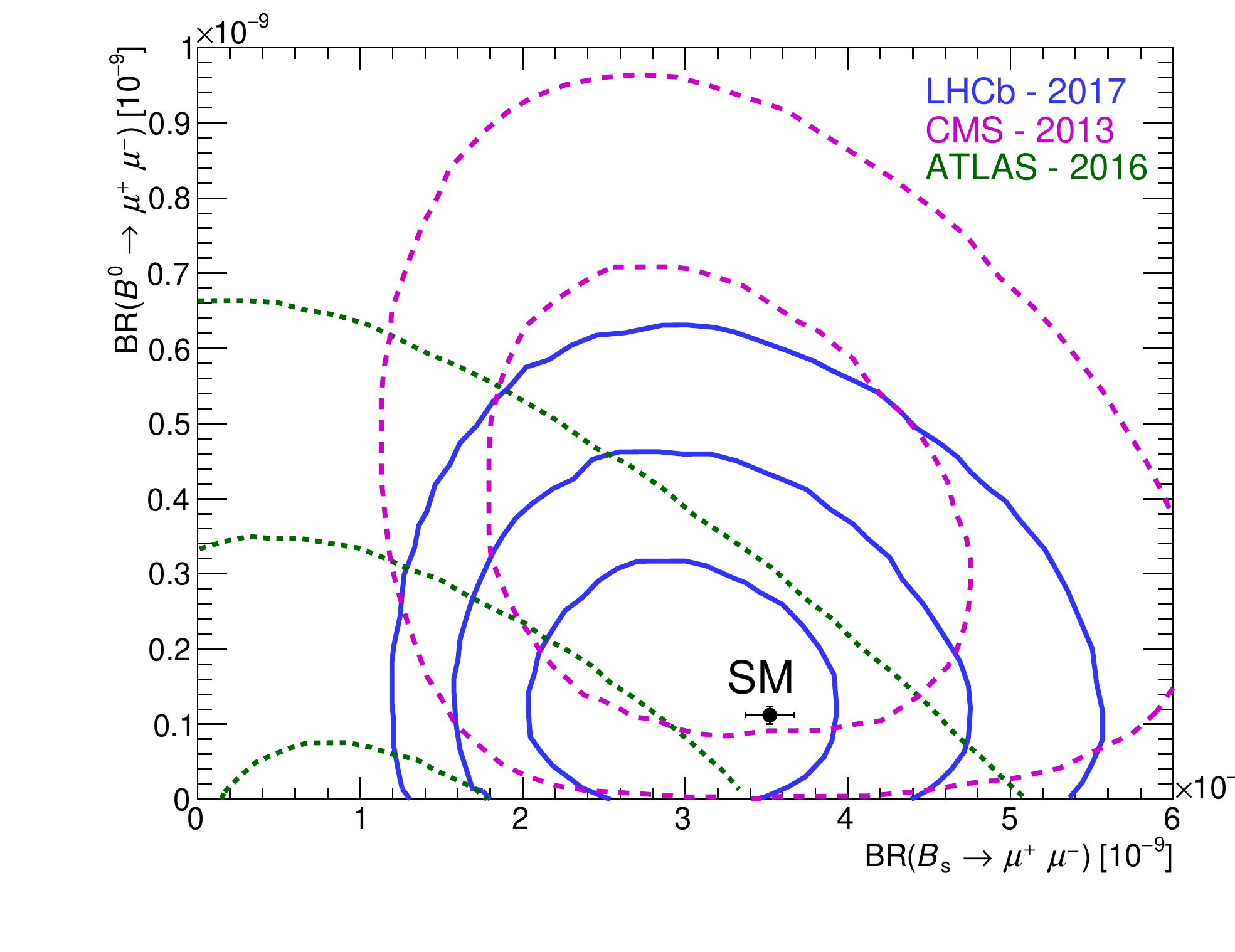}}
     \vspace*{8pt}
     \caption{Lines of constant likelihood at 1, 2, and 3 $\sigma$
       (($-2\Delta \ln \mathcal L = 2.30, 6.18, 11.83$)
       for the three measurements at LHC experiments\cite{lhcb_bsmm,bsmm_cms_obs,bsmm_atlas}.
       The SM predictions with 1 $\sigma$ uncertainty is also shown.\protect\label{fig:bsmm_exps}}
   \end{figure}

  \vskip 2mm
  The current experimental results on the $BR(B^0_s \to \mu^+ \mu^-)$ and $BR(B^0 \to \mu^+ \mu^-)$ from the
  LHC experiments:
  
   \begin{align}
    \overline{\text{BR}}(B^0_s\to \mu^+ \mu^-)_{\rm CMS}      &= (3.0^{+1.0}_{-0.9}) \times 10^{-9}, \\
    \overline{\text{BR}}(B^0_s\to\mu^+\mu^-)_{\rm LHCb \; 2017} &= (3.0 \pm 0.6^{+0.3}_{-0.2})\times 10^{-9}, \\
    \overline{\text{BR}}(B^0_s\to \mu^+ \mu^-)_{\rm ATLAS}     &= (0.9^{+1.1}_{-0.8}) \times 10^{-9}
    \label{eq:bsmumu_exps} 
  \end{align}
   are summarized in Fig.~\ref{fig:bsmm_exps}.
   They are all in good agreement with the SM predictions (Eq.~\ref{eq:bsmumu_SM}).
   
   \vskip 2mm
   The search for the tauonic modes $B^0 \to \tau^+ \tau^-$ and $B^0_s \to \tau^+ \tau^-$ is today of great
   interest in view of the recent hints of lepton flavor non-universality, as will be discussed in Section~\ref{ssec:lfu}.
   Possible explanations for these deviations include
   lepto-quarks~\cite{Beciveric_08501, DeMedeiros_01084}, $W^{'} / Z^{'}$  bosons~\cite{Crivellin_07928}
   and two-Higgs-doublet
   models~\cite{Beciveric_08501, Crivellin_03477} that could also enhance the $BR(B^0_{(s)}\to \tau^+ \tau^-)$
   by several orders of magnitude~\cite{Dighe, Alonso2, Cline}.

   While the branching fractions  of the $B^0_s \to \tau^+ \tau^-$ and $B^0 \to \tau^+ \tau^-$ decay modes
   are larger than those with electrons or muons in the final state due to the reduced helicity suppression,
   the experimental search for these modes is complicated by the presence of at least two undetected neutrinos,
   originating from the decay of the $\tau$ leptons. The BaBar collaboration has searched for the
   $B^0 \to \tau^+ \tau^-$ mode~\cite{BaBar_btautau} and published an upper limit
   $BR(B^0 \to \tau^+ \tau^-)< 4.1 \times 10^{-3}$
   at 90\% CL, while no experimental results were available 
   for the $B_s \to \tau^+ \tau^-$ mode before the LHC era.
   
\vskip 2mm
The LHCb collaboration has searched for these decay modes~\cite{lhcb_btautau} using the Run~1 data set and
reconstructing the $\tau$
leptons through the decay $\tau \to \pi \pi \pi \nu_{\tau}$, which proceeds predominantly through the decay
chain $\tau \to a_1(1260) \nu_{\tau}\; , a_1(1260) \to \rho(770) \pi$.
Assuming no contribution from $B^0 \to \tau^+ \tau^-$ decays, the first upper limit is set
on the $B_s \to \tau^+ \tau^-$ decay mode, $BR(B_s \to \tau^+ \tau^-) <  5.2 \times 10^{-3}$ at 90\% CL.
The analysis was then repeated for the $B^0 \to \tau^+ \tau^-$ decay, by changing the signal model,
and an upper limit on the branching fraction $BR(B^0 \to \tau^+ \tau^-) < 1.6 \times 10^{-3}$ at 90\% CL is set.
This upper limit represents an improvement of a factor 2.6 with respect to the BaBar result
and it is currently the world best upper limit.

\vskip 2mm
LFV decays $B^0 \to e^{\pm} \mu^{\mp}$ and $B^0_s \to e^{\pm} \mu^{\mp}$ are forbidden in the SM and their observation
would be a clear evidence of NP. Their study is particularly interesting today in light of lepton non-universality effects
discussed in the Section~\ref{ssec:lfu}. The LHCb experiment has recently updated the results from a search for LFV decays
$B^0_{(s)} \to e^{\pm } \mu^{\mp}$ using the full Run1 dataset~\cite{lhcb_04111}.
The observed yields are consistent with the background-only hypothesis and upper limits on the branching fractions
are determined to be $BR(B^0_s \to e^{\pm} \mu^{\mp}) < 5.4 \times 10^{-9}$
and $BR(B^0 \to e^{\pm} \mu^{\mp}) < 1.0 \times 10^{-9}$ at 90 \% CL.

  \section{Semi-leptonic decays}
  \label{sec:semi-leptonic}

  Semi-leptonic $b-$decays mediated by neutral currents are based on the $b \to q \ell^+ \ell^-$
  quark-level transition, where  $\ell = e,\mu,\tau$ and $q=s,d$.
  Compared to leptonic or radiative decays, which are only sensitive to effects in $Q^{(')}_{10}$
  or $Q^{(')}_7$, respectively, semi-leptonic decays provide important insight into the SM structure
  as they are sensitive to NP contributions
  to the operators $Q^{(')}_7$, $Q^{(')}_9$
  and $Q^{(')}_{10}$ through a diverse and broad set of observables,
  as branching fractions, ratios of branching fractions and angular distributions.

  \vskip 2mm
  Unfortunately in this case, the hadronic uncertainties are more challenging than for purely
  leptonic decay modes as the lepton pair can also originate from a photon that can enhance
  the decay rate by orders of magnitude when the dilepton invariant mass squared $q^2$ is close
  to the mass of charmonium resonances $J/\psi(1S)$ and $\psi(2S)$.
  In addition, exclusive decays $B \to M \ell^+ \ell^-$  to a meson $M$ require the knowledge of the
  $B \to M$  form factors in the full kinematic range   $4 m^2_{\ell} < q^2 < (m_B-m_M)^2$.

  \vskip 2mm
  From the experimental point of view these decays are challenging because of the small
  branching fraction ($ o(10^{-6})$ for $b \to s \ell^+ \ell^-$ and $o(10^{-8})$
  for $b \to d \ell^+ \ell^-$ transitions) and for the presence of low-$p_{\rm T}$ electrons and muons
  in the final state, which are more difficult to reconstruct, in particular in a hadronic environment.

  \vskip 2mm
  Current data on exclusive decays based on $b \to s \ell^+ \ell^-$ transitions seem to show a coherent
  pattern of deviations from SM predictions in the values of branching
  fractions~\cite{lhcb_BR_BKst, lhcb_BR_BKst_2, lhcb_BR_BPhi, lhcb_BR_BLambda},
  shapes of angular distributions~\cite{lhcb_P5, belle_P5} and in tests of lepton flavour
  universality (LFU)~\cite{lhcb_RK, lhcb_RKstar}, as will be discussed in the following.

  \vskip 2mm
  Unfortunately, a conclusive statement about possible beyond-SM effects can not be drawn
  for all the decays that are affected by hadronic uncertainties, as these uncertainties
  could be the origin of the measured discrepancies~\cite{Jager,Descotes,Lyon,Capdevila, Chobanova, Ciuchini}.
  
  Ratios of branching fractions involving different lepton flavours~\cite{Hiller2, Hiller3}
  are instead practically free of hadronic uncertainties and are of particular interest
  as they can be used to test the LFU in the SM.
  The intriguing hints of LFU deviations that have been recently observed in B semi-leptonic transitions
  are therefore currently at the center of a lively discussion within the flavor community. 

  \subsection{Branching fractions}
  \label{ssec:semi-leptonic-br}
      
\begin{figure}[htb]
        \centerline{\includegraphics[width=0.65\linewidth]{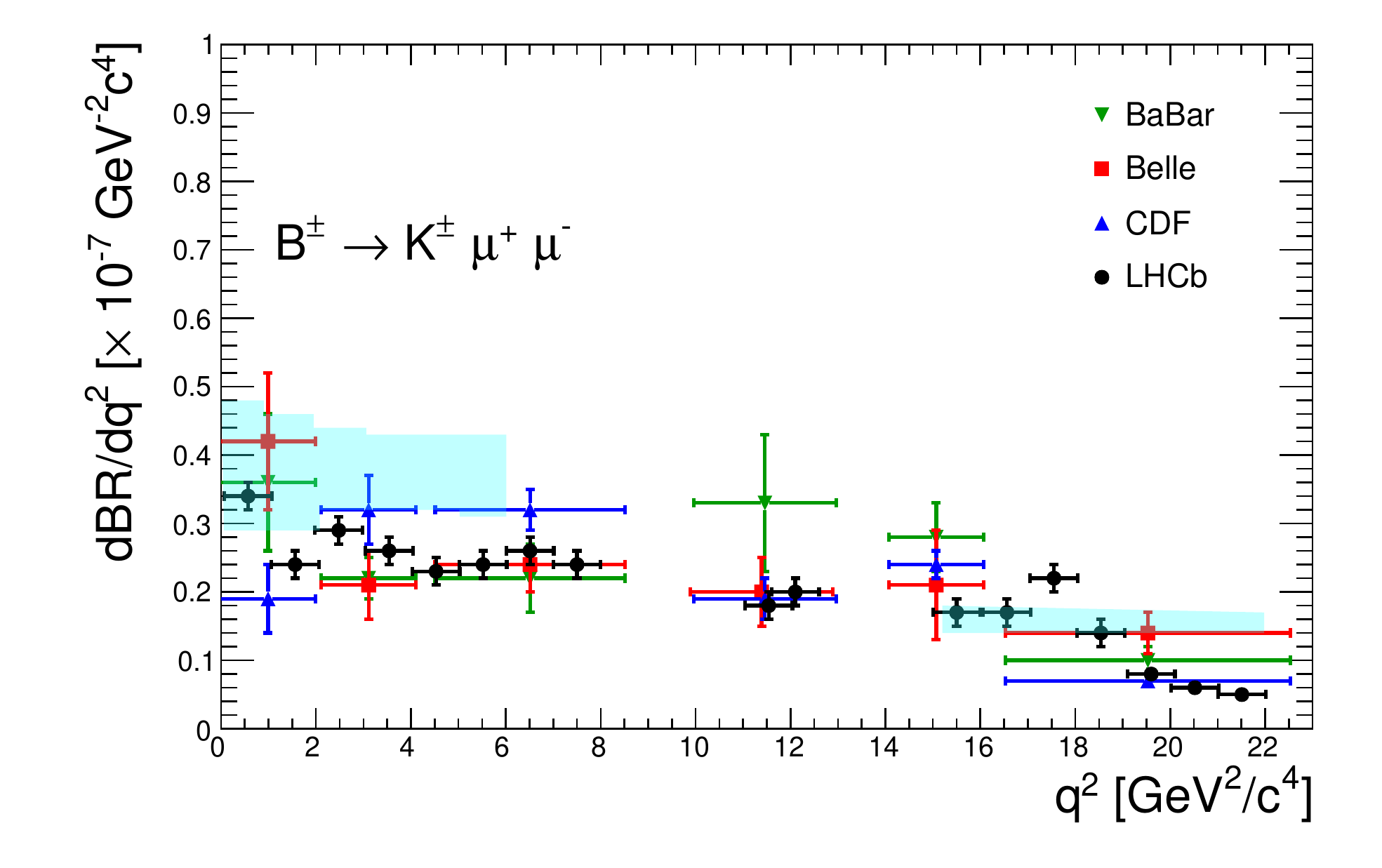}}
        \centerline{\includegraphics[width=0.65\linewidth]{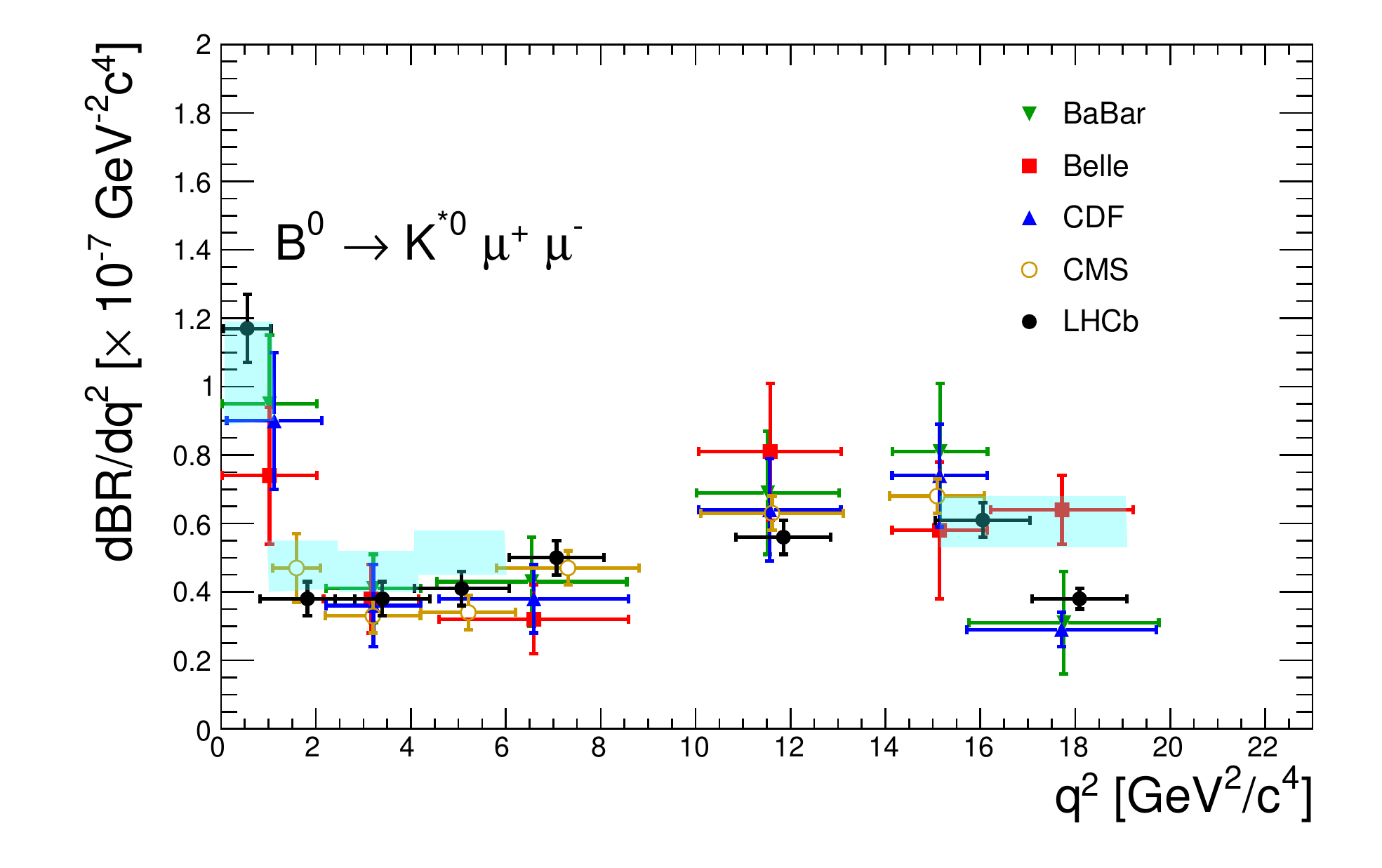}}
        \vspace*{8pt}
        \caption{Differential branching fractions as a function of the di-muon invariant mass squared $q^2$, for
          several $b \to s \ell \ell$ decay modes measured by the BaBar\cite{babar_BR_bsll},
          Belle~\cite{belle_BR_bsll}, CDF\cite{cdf_BR_bsll}, CMS~\cite{cms_BR_bsll} and LHCb~\cite{lhcb_BR_BKst, lhcb_BR_BKst_2}
          experiments.
          The point are experimental measurements while the shaded regions are theoretical predictions.
          For details see text.\protect\label{fig:br_bsll_1}}
      \end{figure}

\begin{figure}[htb]
        \centerline{\includegraphics[width=0.65\linewidth]{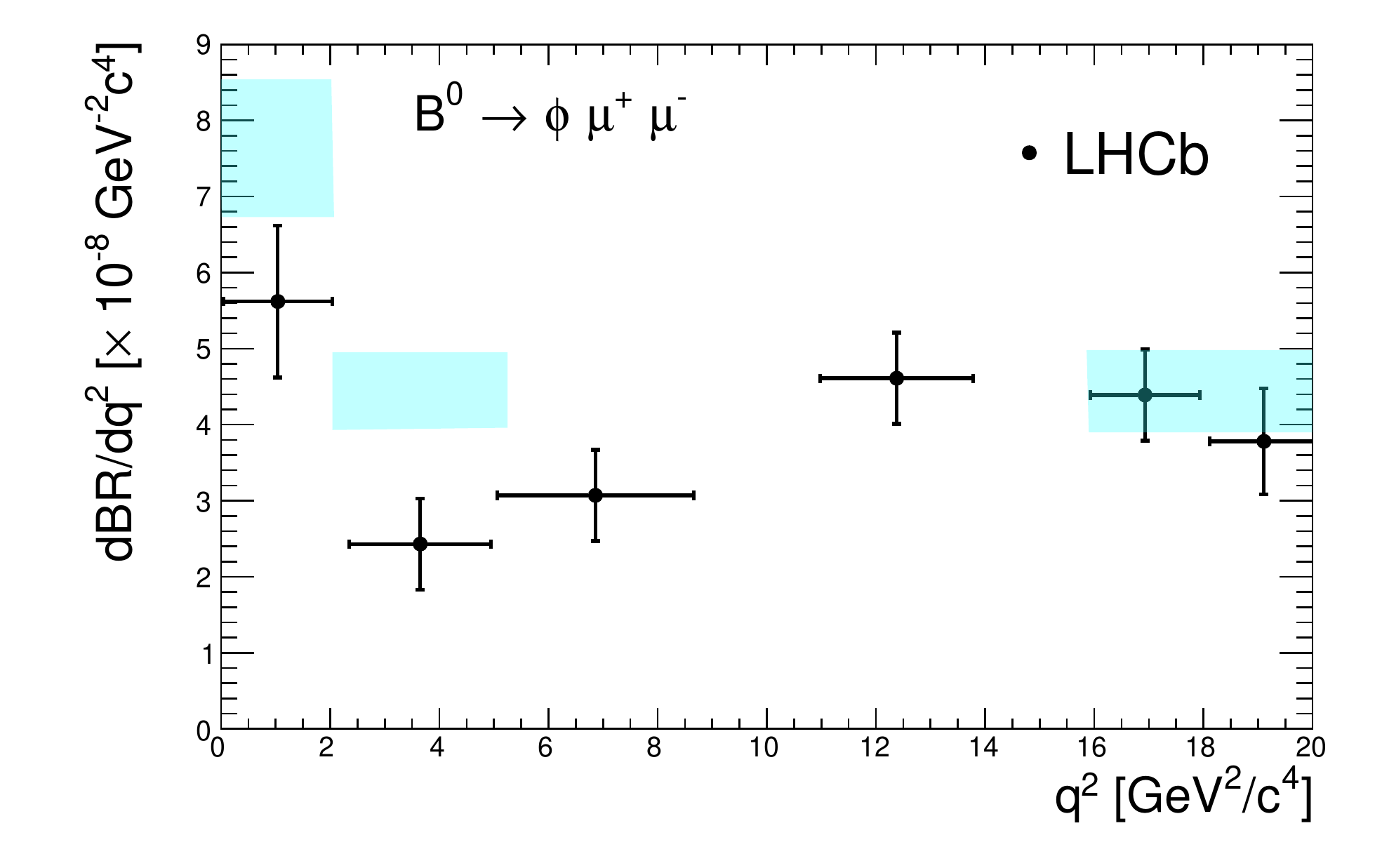}}
        \centerline{\includegraphics[width=0.65\linewidth]{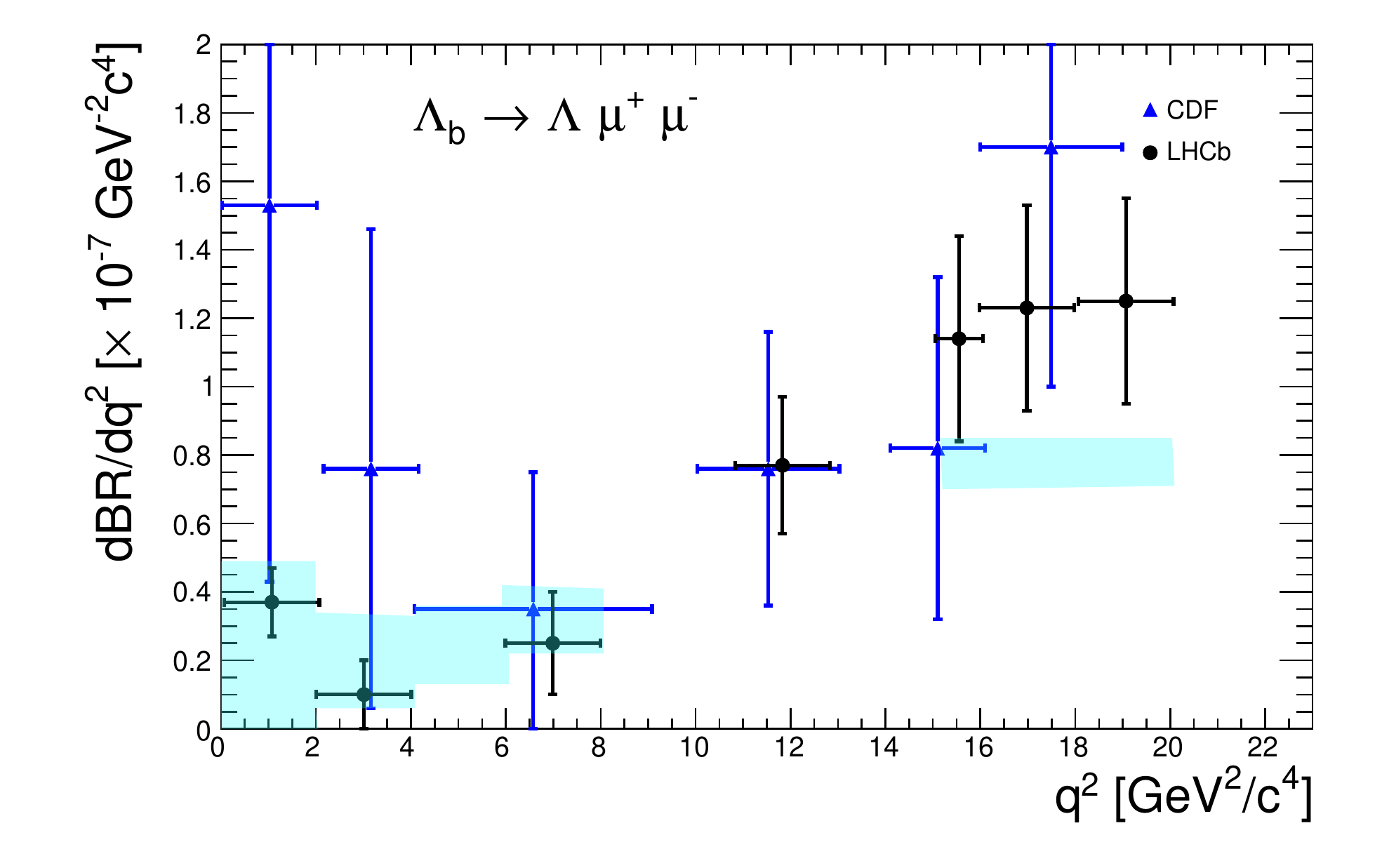}}
        \vspace*{8pt}
        \caption{Differential branching fractions as a function of the di-muon invariant mass squared $q^2$, for
          several $b \to s \ell \ell$ decay modes measured by CDF~\cite{cms_BR_bsll} and
          LHCb\cite{lhcb_BR_BPhi, lhcb_BR_BLambda} experiments.
          The point are experimental measurements while the shaded regions are theoretical predictions.
          For details see text.\protect\label{fig:br_bsll_2}}
      \end{figure}

      A summary of experimental measurements of the differential branching fractions for the 
      $b \to s \ell^+ \ell^-$  quark-mediated processes as $B^+ \to K^+ \ell^+ \ell^-$,
      $B^0 \to K^* \ell^+ \ell^-$, $B_s \to \phi \ell^+ \ell^-$, and $\Lambda_b \to \Lambda \ell^+ \ell^-$,
      as a function of $q^2$ , is provided in Fig.~\ref{fig:br_bsll_1} and Fig.~\ref{fig:br_bsll_2}.

      The results obtained from BaBar~\cite{babar_BR_bsll}, Belle~\cite{belle_BR_bsll},
      CDF~\cite{cdf_BR_bsll},  CMS~\cite{cms_BR_bsll},  and LHCb~\cite{lhcb_BR_BKst, lhcb_BR_BKst_2, lhcb_BR_BPhi, lhcb_BR_BLambda},     
      experiments are now much more precise than
      the corresponding SM predictions.
      The theoretical predictions in Fig.~\ref{fig:br_bsll_1} and Fig.~\ref{fig:br_bsll_2}
      mostly use LCSR predictions for the form-factors at large recoil (low $q^2$)\footnote{
      See, for example, Refs.~\refcite{Ball},\refcite{Bharucha},\refcite{Sentitemsu} for $B \to P$ form factors,
      with $P=$ pseudo-scalar particle, and Refs.~\refcite{Bharucha_2},\refcite{Ball_2} for $B \to V$ form factors,
      with $V$=vector particle.}
      and lattice QCD for those at low recoil (large $q^2$)~\footnote{See Ref.~\refcite{Bailey} for $B \to P$ 
      transitions and Refs.~\refcite{Horgan},~\refcite{Horgan2} for $B \to V$ transitions.}.

      In general, the experimental measurements of the branching fractions tend to lie below the
      SM expectations across the full $q^2$ range. The discrepancy is largest for the
      $B_s \to \phi \mu^+ \mu^-$ decay in the large recoil region  ($1<q^2<6$ GeV$^2$/c$^4$) where the experimental
      points are more than $3 \,\sigma$ away from the SM predictions.
      The exception to this trend is the differential branching fraction of the $\Lambda_b \to \Lambda \mu^+ \mu^-$ where,
      at least at low recoil, the measured branching fraction is above (but consistent with) the SM predictions.
      
      \subsection{Angular distributions}
      \label{ssec:angular_distributions}

      NP contributions can also modify angular distributions of $B \to K \ell^+ \ell^-$
      (with $B=B^+,B^0$ and $K=K^+, \overline{K}^0$) and $B^0 \to K^{0*} \ell^+ \ell^-$ decays.
      The angular distribution of $B^+ \to K^+ \ell^+ \ell^-$ decays is described only by one angle, $\theta_{\ell}$,
      which is the angle between the flight direction of the $\ell^+$ and the direction of the $B$ in the di-lepton rest-frame.
      The differential decay rate is given by:

      \begin{equation}
        {{\rm d}^2 \Gamma(B \to K \ell^+ \ell^-)
          \over
              {\rm d} \cos \theta_{\ell} {\rm dq}^2}
        = {3 \over 4} (1-F_{\rm H}) (1-\cos^2 \theta_{\ell}) + {1 \over 2} F_{\rm H} + A_{\rm FB} \cos \theta_{\ell}
        \label{eq:ang_BKll}
      \end{equation}

      where both $F_{\rm H}$ and the forward-backward asymmetry $A_{\rm FB}$, depend on $q^2$.
      In SM both $F_{\rm H}$ and $A_{\rm FB}$ are tiny, but in presence of NP can be enhanced.
      LHCb has measured $F_{\rm H}$ and $A_{\rm FB}$ in $B^+ \to K^+ \mu^+ \mu^-$ decays~\cite{lhcb_BKll_angular} and
      found them consistent with SM expectations. Recently~\cite{cms_AFB} CMS has performed an angular analysis of the
      $B^+ \to K^+ \mu^+ \mu^-$ mode based on an integrated luminosity of 20.5 fb$^{-1}$ collected in Run 1.
      They measured values $A_{\rm FB} = -0.14^{+0.07}_{-0.06} \pm 0.03$ and $F_{\rm H} = 0.38^{+0.17}_{-0.21} \pm 0.09$ in the
      $q^2$ range $[1,6.0]$ GeV$^2$/c$^4$ are both compatible with the SM predictions.

      The angular distribution of $B^0 \to  K^{0*} \ell^+ \ell^-$ decays is more involved as it is defined by
      two additional angles, the angle $\theta_K$ between the direction of the kaon and the direction
      of the $B$ in the $K^{0*}$ rest frame and the angle $\phi$  between the decay plane of the two leptons
      and the $K^*$ in the rest frame
      of the $B$\footnote{For a discussion about angular conventions see Ref.~\refcite{Gratrex}.}.      
      The $B^0 \to K^{0*} \mu^+ \mu^-$ angular distribution can be described by eight angular observables:
      $F_{\rm L}$ (the longitudinal polarization of the $K^*$), $A_{\rm FB}$, and six additional observables
      that vanish when integrated over $\phi$\cite{rarebdecays}.
      
\begin{figure}[h]
        \centerline{\includegraphics[width=0.65\linewidth]{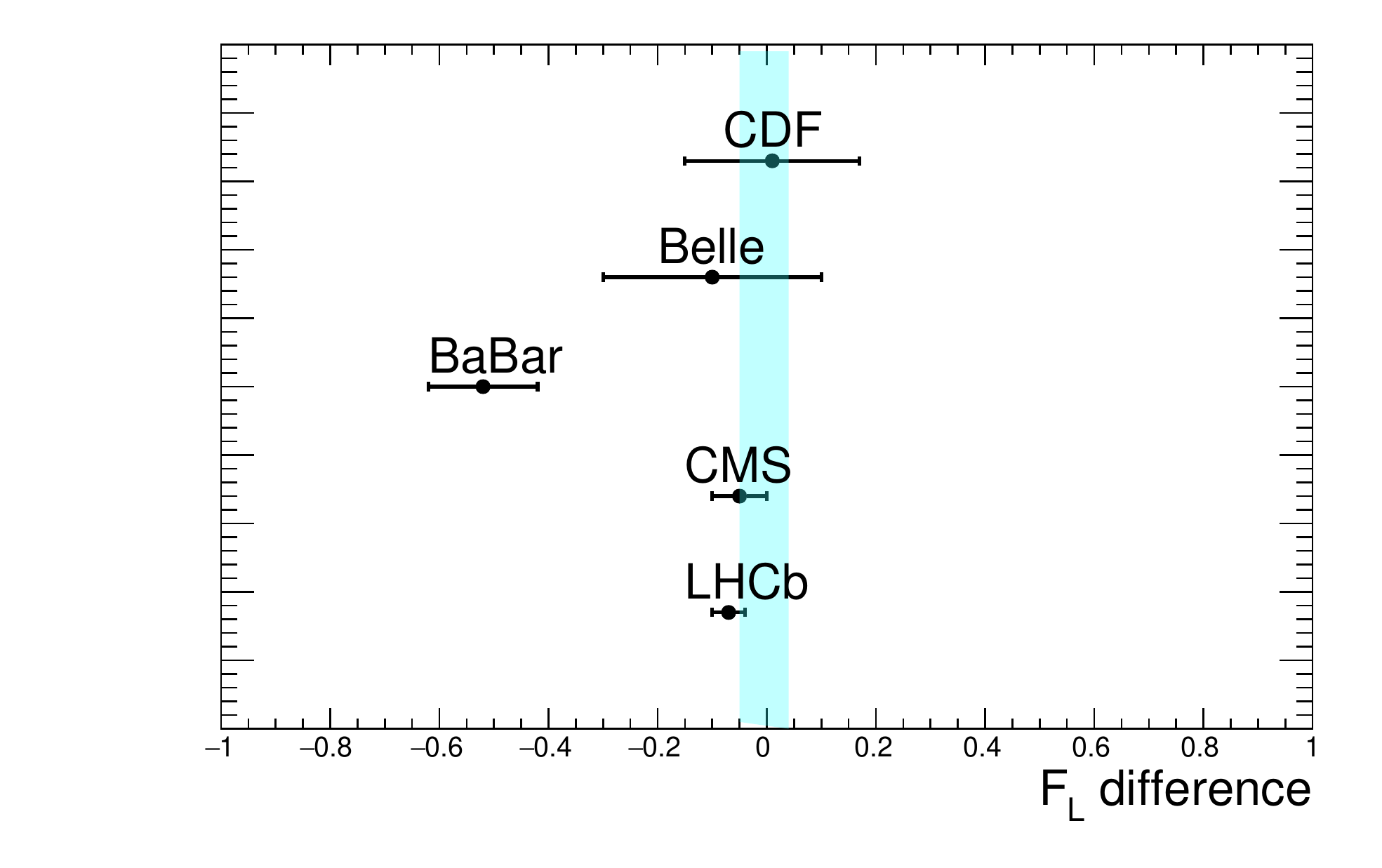}}
        \centerline{\includegraphics[width=0.65\linewidth]{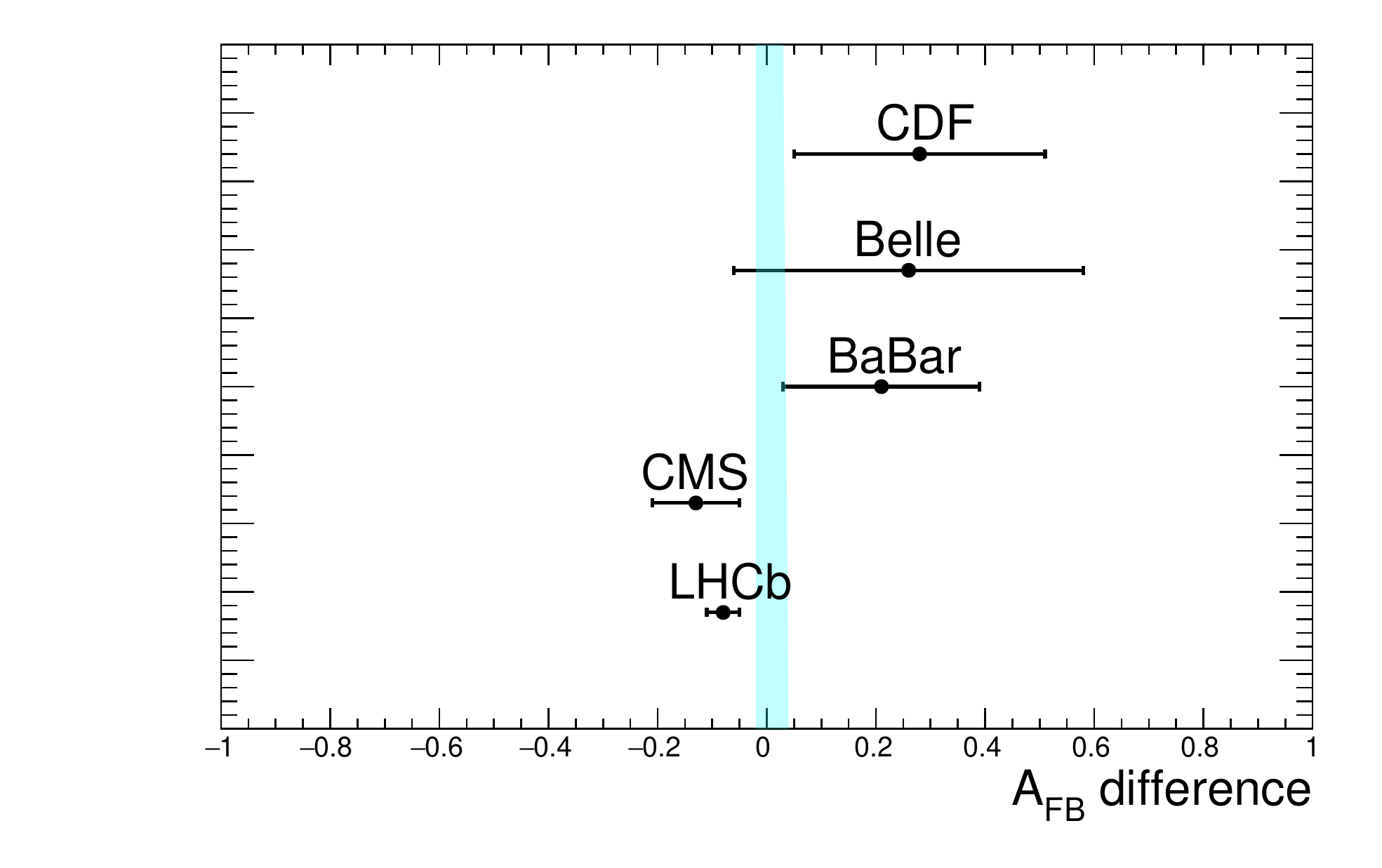}}
        \vspace*{8pt}
        \caption{Fraction of longitudinal polarisation $F_{\rm L}$ of the $K^{0*}$ system,
          and dilepton system forward-backward asymmetry $A_{\rm FB}$ measured by the BaBar~\cite{babar_AFB},
          Belle~\cite{belle_AFB}, CDF~\cite{cdf_AFB}, CMS~\cite{cms_AFB} and LHCb~\cite{lhcb_AFB}
          collaborations in the dimuon mass squared range $1 < q^2 < 6$ GeV$^2$.
          The Figure is reproduced from Ref.~\cite{rarebdecays}.\protect\label{fig:FL-AFB}}
      \end{figure}

      The current measurements of $A_{\rm FB}$ and $F_{\rm L}$ from BaBar~\cite{babar_AFB}
      Belle~\cite{belle_AFB}, CDF\cite{cdf_AFB}, CMS\cite{cms_AFB_Kst} and
      LHCb~\cite{lhcb_AFB} experiments  are shown in Fig.~\ref{fig:FL-AFB}, 
      along with SM predictions~\cite{theory_AFB, theory_AFB_2}:     
      they are consistent with each other and also with the SM predictions, with the larger tension observed
      in the BaBar measurement. 

\begin{figure}[h]
        \centerline{\includegraphics[width=0.8\linewidth]{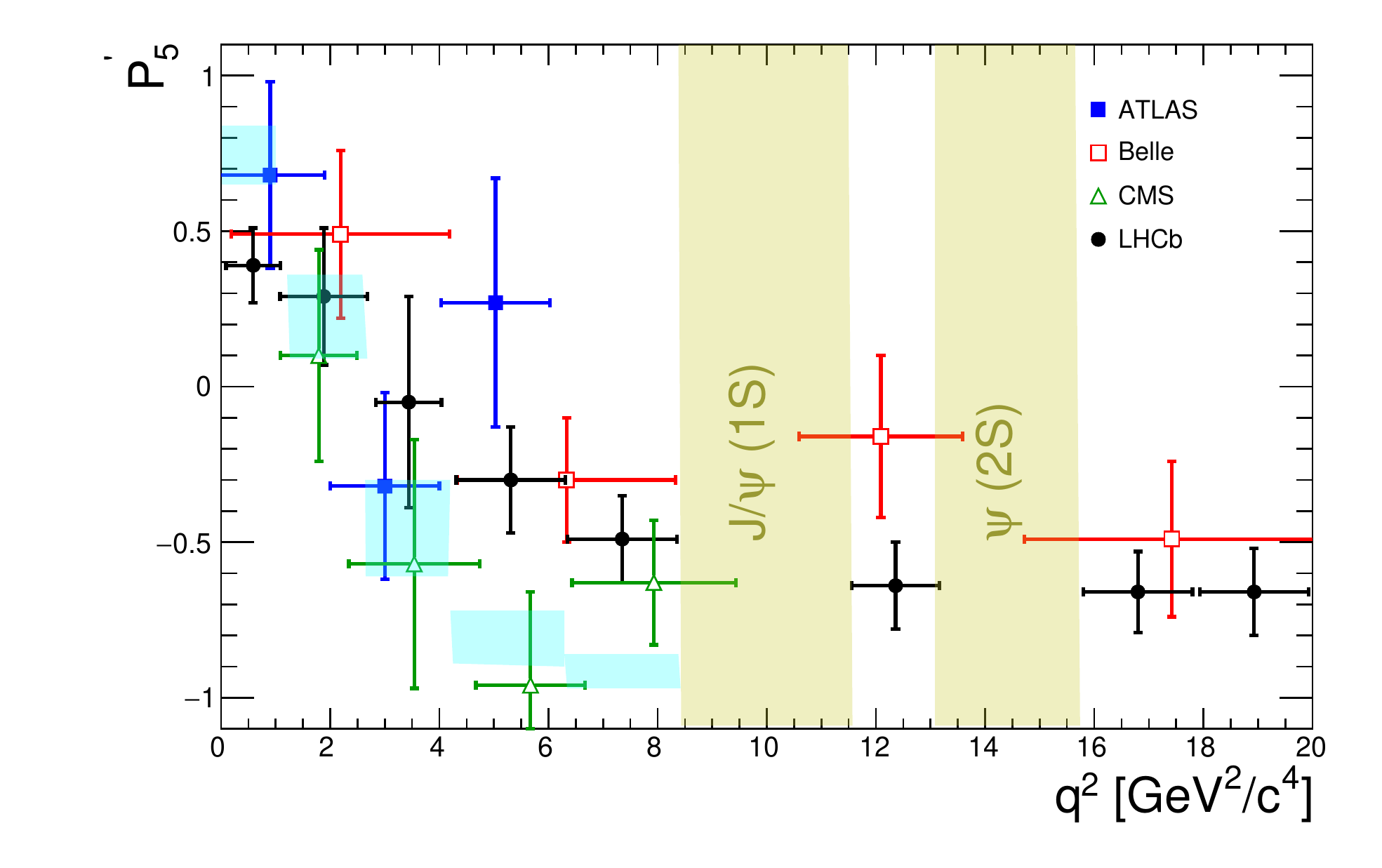}}
        \vspace*{8pt}
        \caption{Measurements of the $P^{'}_5$ observable by LHCb~\cite{lhcb_P5}, Belle~\cite{belle_P5},
          ATLAS~\cite{ATLAS_angular} and CMS~\cite{CMS_angular} as a function
          of the di-muon invariant mass squared $q^2$. Theoretical predictions~\cite{P5_DHMV}
        are shown a shaded areas. The regions near to the $J/\psi$ and $\psi(2S)$ resonances are excluded as
        the amplitude in these mass regions has large theoretical uncertainty.
        \protect\label{fig:P5}}
      \end{figure}

      ATLAS~\cite{ATLAS_angular}, CMS~\cite{CMS_angular}, LHCb~\cite{lhcb_P5} 
      and Belle~\cite{belle_P5} have also performed a full angular analysis that is sensitive
      to the observables that depend on the $\phi$ angle. While the large majority of these additional
      observables is consistent with the SM expectations, the LHCb measurement of the $P^{'}_5$ variable shows hints of deviations
      from SM predictions~\cite{P5_DHMV} in the $q^2$ region $4 < q^2 < 8$ GeV$^{2}$/c$^{4}$, as shown in Fig.~\ref{fig:P5}.
      The CMS measurement of the same variable, however, is in agreement with the expectations.
      
      While the accuracy on the experimental points is currently statistically limited,
      and is expected to improve with the increase of the datasets, the theory predictions
      suffer for hadronic uncertainties due to non factorisable corrections.
      The size of those corrections and the theory methods required to compute them vary strongly with $q^2$.
      The main challenge in exclusive $b \to s \ell^+ \ell^-$ decays at low $q^2$ is represented by soft gluon corrections
      to the charm loop, that have been estimated in LCSR~\cite{Khodjamirian_4945, Khodjamirian_0234} but remain the
      largest source of uncertainty. There are also studies~\cite{Ciuchini} that claim that no deviation is present once all the theoretical
      uncertainties are taken into account.

      \vskip 2mm
      If the observed discrepancies are due to long-distance charm loop contributions,
      these can interfere with short-distance effects of the $J/\psi$ and $\psi(2S)$ resonances
      and affect the $q^2$ region far from the pole mass regions.
      To model properly these contributions, LHCb has recently performed a study~\cite{lhcb_long_distance}
      of the $B^+ \to K^+ \mu^+ \mu^-$
      decay channel, measuring for the first time the phase difference between the short-distance and
      narrow-resonance amplitudes by analysing the dimuon invariant mass spectrum
      including the $J/\psi$ and $\psi(2S)$ resonances.
      The analysis is based on the Run 1 dataset.
      The long-distance contribution to the $B^+ \to K^+ \mu^+ \mu^-$ decay is modelled as a sum of
      relativistic Breit-Wigner amplitudes representing different vector meson resonances
      decaying to muon pairs, each with their own magnitude and phase.

      The measured phases of the $J/\psi$ and $\psi(2S)$ resonances are such that the interference with the
      short-distance component in di-muon mass regions far from their pole masses is small.
      Unfortunately, the limited dataset does not allow so far to resolve the four
      four-fold ambiguity $(\delta_{J/\psi}|\delta_{\psi (2S)} | 0,\pi | 0, \pi)$.
      This ambiguity can be removed with a larger dataset in the future.

      \vskip 2mm
      Understanding the role of the charm-loops will remain a mandatory task before physics beyond the
      SM can be considered the responsible of the angular anomalies of the $B^0 \to K^{0*} \mu^+  \mu^-$ mode and branching
      fractions deviations. Progress on this topic will require from one side to compute charm-loop contributions
      consistently using a single approach, and on the other side to constrain hadronic uncertainties by extracting
      relevant information with more refined measurements.

      \subsection{Lepton flavor universality tests}
      \label{ssec:lfu}

      Lepton Flavor Universality (LFU) is not a fundamental symmetry of the SM:
      it is accidental in the gauge sector, where the gauge bosons  equally couple to the different lepton flavors.
      LFU is broken by the Higgs couplings to masses, which are flavor specific, but they have
      a negligible effect on the partial widths of the decays.
      Ratios of the partial widths which involve different lepton flavors,
      referred to as $R$-ratios, are therefore expected to be unity up to
      corrections from phase-space differences due to the different masses of the leptons.

      In the recent years several hints of LFU in semileptonic $B$ decays have been reported and are currently
      at the centre of a lively discussion within the flavor community.      
      The observed LFU violations in $B$ semileptonic decays can be classified into two categories,
      following the underlying quark-level transitions:
      
      \begin{itemize}      
      \item[-] $\tau/\mu$ (and $\tau/e$) LFU violations in $b \to c \ell \overline{\nu}$ charged
        currents;
      \item[-] $\mu/e$ LFU universality in $ b \to s \ell \overline{\ell}$ neutral currents.
      \end{itemize}

      LFU violations in charged currents have been observed in 
      $R(D^{(*)})=BR( B \to D^{(*)} \tau \nu_{\tau} / BR(B^{(*)} \to D^{(*)} \ell \nu_{\ell}$, with $\ell = e, \mu$
      by several
      experiments~\cite{BaBar_RD_RDst_0,BaBar_RD_RDst,Belle_RD_03233,Belle_RDst_07923,Belle_RD,lhcb_RDst,lhcb_RDst_2}.
      The HFLAV average values~\cite{HFLAV}:
      
      \begin{align}
        R(D^{*}) & = 0.304 \pm 0.013 \pm 0.007 \,\\
        R(D) = & = 0.407 \pm 0.039 \pm 0.024.
        \label{eq:RDst_RD}
      \end{align} 
      deviate from the SM predictions, $R(D^*) = 0.252 \pm 0.003$ ~\cite{Fajfer_Kamenik}
      and $R(D) = 0.300 \pm 0.008$ ~\cite{FLAG_RD}
      by $3.4 \,\sigma$ and $2.3 \,\sigma$, respectively.
      A combined fit which takes properly into account the correlations between the two measurements
      gives $4.1 \;\sigma$ deviation
      from the SM predictions, with a NP contribution which is about 10\% of the SM amplitude.

      LFU violation in neutral currents has been measured in
      $b \to  s \ell^+ \ell^-$ (with $\ell = e, \mu$)
      transitions involving ratios of branching fractions such as~\cite{Hiller2, Hiller3}
      
      \begin{align}
        R_K & = { BR(B^+ \to K^+ \mu^+ \mu^-)  \over BR(B \to K^+ e^+ e^-) }\\
        R_{K^*} & = {BR (B^0 \to K^{0*} \mu^+ \mu^-) \over BR(B^{0} \to K^{*0} e^+ e^-)}
        \label{rk_rkst}
      \end{align}
      
      These ratios are extremely interesting observables as they are very well predicted in the SM to be
      $R_K = R_{K^*} = 1$ in a large range of $q^2$, with a theoretical uncertainty of o(1\%)~\cite{Bordone_07633}.

\begin{figure}[h]
        \centerline{\includegraphics[width=0.8\linewidth]{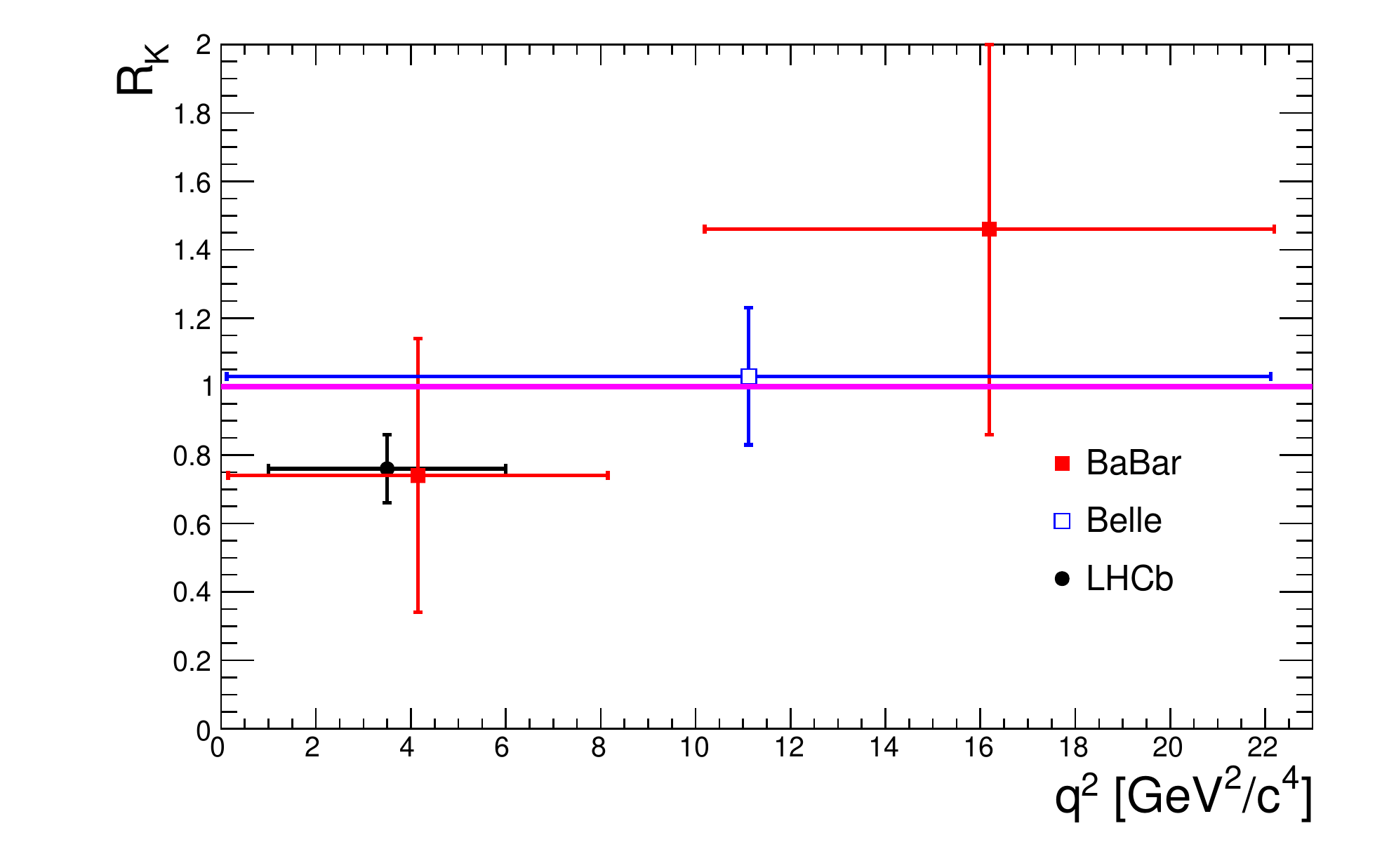}}
        \vspace*{8pt}
        \caption{Measurements of the ratio $R_K = BR(B^+ \to K^+ \mu^+ \mu^-)/BR(B^+ \to K^+ e^+ e^-)$ from
          B-factories experiments~\cite{Babar_RK, Belle_RK} and LHCb~\cite{lhcb_RK}.
          \protect\label{fig:RK}}
      \end{figure}

      Measurements of $R_K$ performed at $e^+ e^-$ colliders operating at the $\Upsilon(4S)$ resonance show values consistent
      with unity with a precision of 20-50\%~\cite{Babar_RK, Belle_RK}.
      LHCb has measured $R_K$ with $\sim {3}$ fb$^{-1}$ of data collected during Run 1~\cite{lhcb_RK}.
      The value of $R_K$ has been measured in the $q^2$ range $1 < q^2 < 6$ GeV$^{2}$/c$^{4}$ and the result,
      $R_K = 0.745^{+0.090}_{-0.074} ({\rm stat}) \pm  0.036 ({\rm syst})$, is shown in Fig.~\ref{fig:RK}.
      This measurement is consistent with the SM predictions at $2.6 \, \sigma$ level.

      \vskip 2mm
      The $q^2$ range used to perform the measurement
      is chosen in such a way to exclude the $J/\psi$ and $\psi(2S)$ resonant regions
      and the high $q^2$ part above the $\psi(2S)$ which includes broad charmonium
      resonances~\cite{lhcb_broad_resonances}.
      In order to minimize possible sources of systematic uncertainties,
      each $B^+ \to K^+ \ell^+ \ell^-$,
      mode is normalized to the corresponding resonant mode, $B^+ \to J/\psi (\ell^+ \ell^-) K^+$.
      The LHCb result is the more precise measurement of the $R_K$ observable so far and its uncertainty is
      statistically dominated.

\begin{figure}[h]
        \centerline{\includegraphics[width=0.65\linewidth]{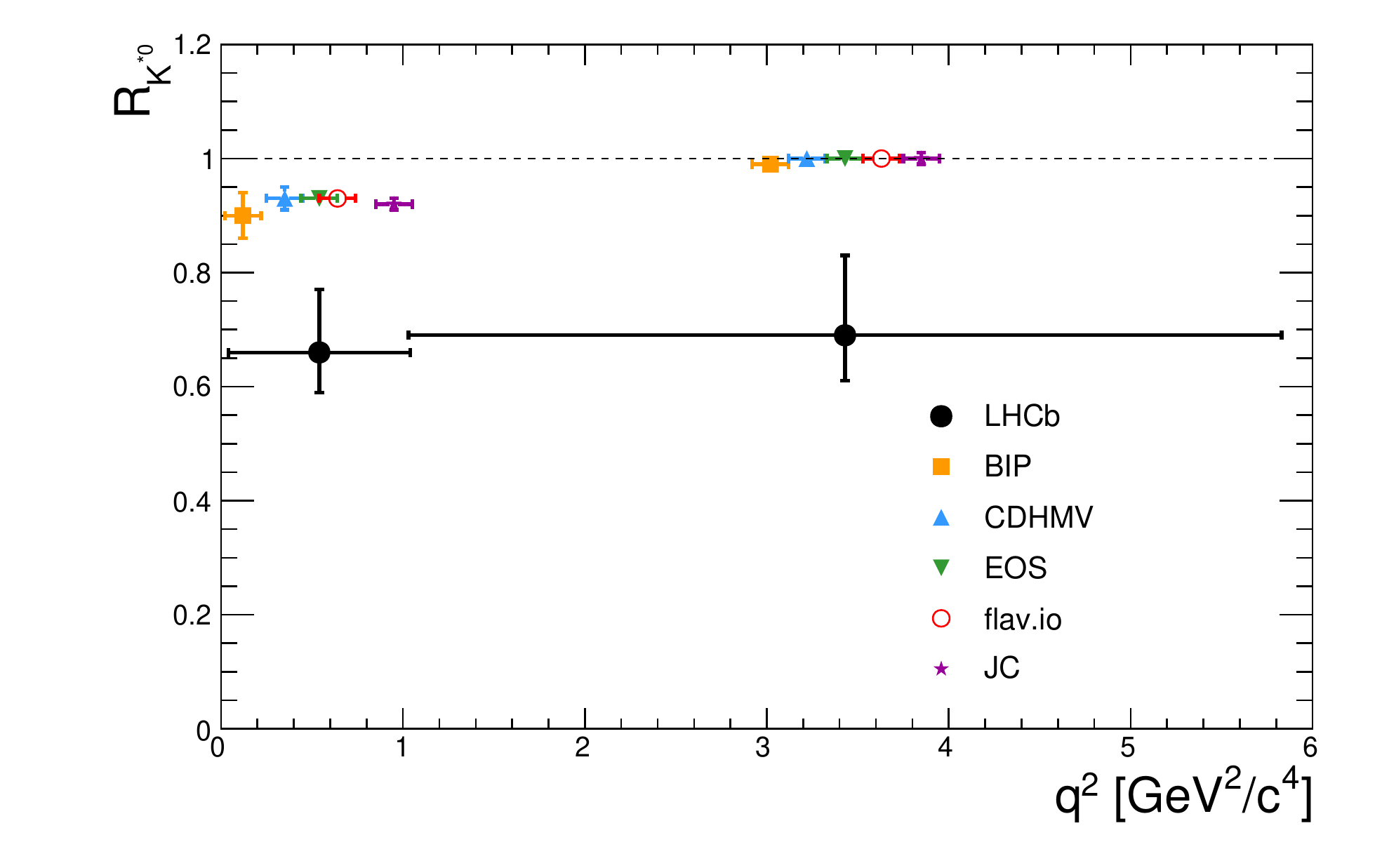}}
        \centerline{\includegraphics[width=0.65\linewidth]{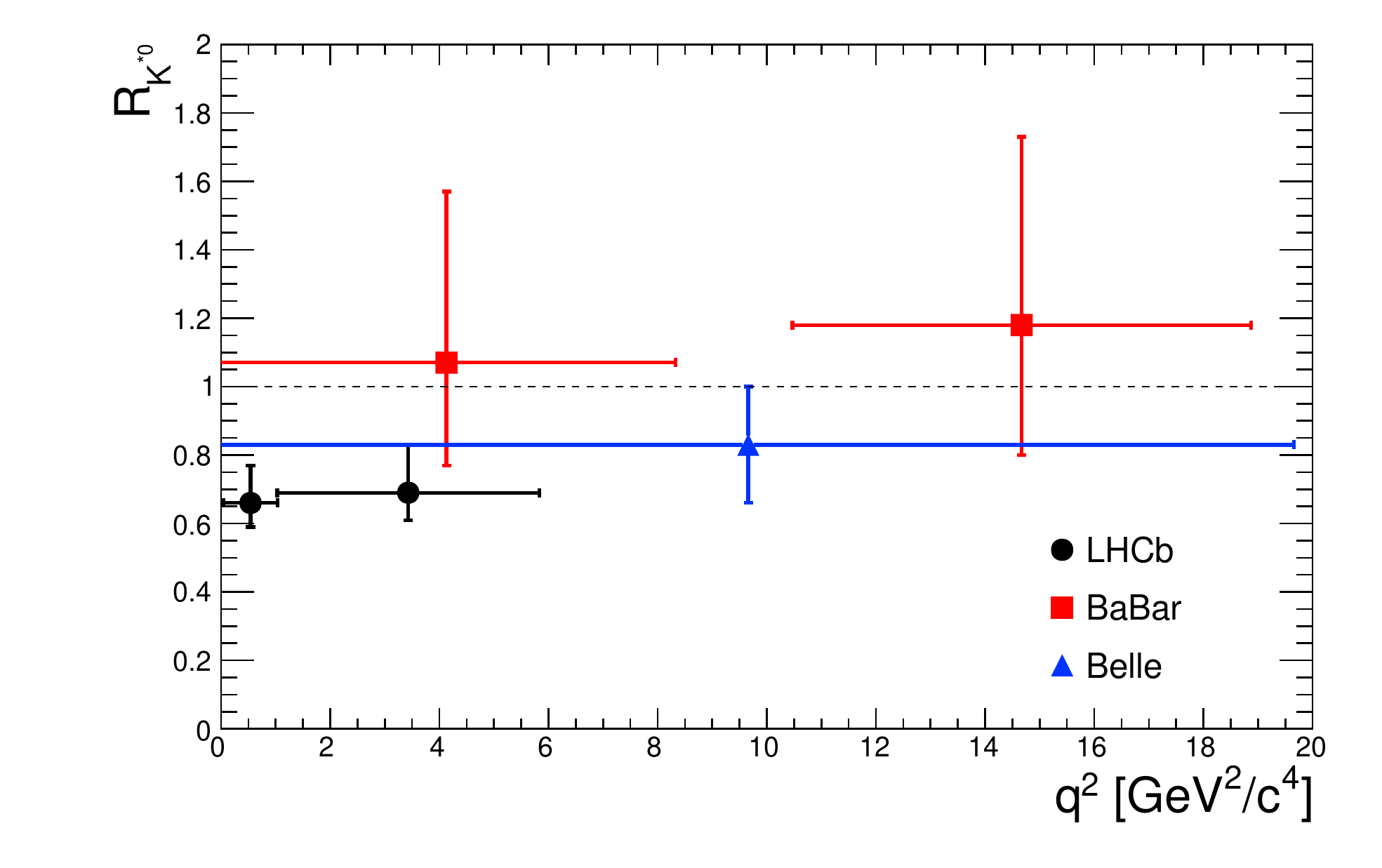}}
        \vspace*{8pt}
        \caption{
          Top: comparison of the LHCb $R(K^{0*})$ measurements with the SM theoretical predictions:
          BIP~\cite{Bordone_07633}
          CDHMV~\cite{Descotes_04239,Capdevila_03156,Capdevila_08672},
          \texttt{EOS}~\cite{Serra_08761,EOS},
          \texttt{flav.io}~\cite{theory_AFB_2,Altmannshofer_09189,flavio} and JC~\cite{Jager_3183}.
          The predictions are displaced horizontally for comparison with the experimental result.
          Bottom: comparison of the LHCb $R(K^{0*})$ measurements with previous experimental results from BaBar~\cite{babar_BR_bsll}
          and Belle~\cite{belle_BR_bsll}.
          Figure reproduced from Ref.~\cite{lhcb_RKstar}.
          \protect\label{fig:RKst}}
      \end{figure}

      Recently~\cite{lhcb_RKstar} LHCb has also measured the ratio of the branching fractions 
      $R_{K^{0*}}$ in two  different $q^2$ ranges,  $0.045 < q^2 < 1.1$ GeV$^2$/c$^4$ and $1.1 < q^2< 1.6$ GeV$^2$/c$^4$.
      The results:

      \begin{align}
        R_{K^{0*}} (0.045 < q^2 < 1.1 {\;\rm GeV}^2/{\rm c}^4)  & = 0.660^{+0.110}_{-0.070} \pm 0.024 \\
        R_{K^{0*}} (1.1 < q^2 < 6.0 {\;\rm GeV}^2/{\rm c}^4)  & = 0.685^{+0.113}_{-0.069} \pm 0.047
        \label{eq:lhcb_RKst}
      \end{align}
 
      are shown in Fig.~\ref{fig:RKst} along with previous measurement from BaBar~\cite{Babar_RK}
      and Belle~\cite{belle_BR_bsll} collaborations and several theory predictions.
      The LHCb results, which are the most precise measurements of $R_{K^{0*}}$ to date,
      are compatible with the Standard Model expectations at the level of $2.1-2.3$ and $2.4-2.5$
      standard deviations in the two $q^2$ regions, respectively. 

      \section{Global fits and New Physics interpretation}
      \label{sec:global_fits}

      As shown in the previous Sections, experimental results on branching fractions, 
      angular distributions, 
      and ratios of branching fractions 
      show deviations from the SM predictions.
      Although individually these results do not show sufficient evidence for NP yet, 
      the pattern that is emerging is certainly interesting as they are all based on the
      same $b \to s \ell^+ \ell^-$ quark level transition.
      
      Several correlated analyses of these anomalies have been performed in order to identify a
      possible universal NP contribution that could relax the tensions observed in the fit and
      provide a coherent description of the data.
      
      All the exclusive FCNC semi-leptonic decays are sensitive to the Wilson
      coefficients $C^{(')}_{(7,9,10)}$, while the leptonic decays $B^{0}_{(s)} \to  \ell^+ \ell^-$  are only sensitive to
      $C^{(')}_{10}$.
      The results obtained by different combined fits of
      $C^{(')}_{7,9,10}$~\cite{theory_AFB,Descotes_04239,Descotes_5683, Altmannshofer_1501,Hurth_00865}
      are in agreement with each other and point towards a large NP contribution in $C_9$ Wilson coefficient.
      
      This is shown in Table~\ref{tab:C} that contains selected results on NP contribution to $C^{(')}_{7,9,10}$
      from Ref.~\refcite{Descotes_04239}.

      \begin{table}[htb]
        \tbl{One-parameter fit allowing for NP only in one of the Wilson coefficients $C^{(')}_{7,9,10}$.
          Best fit point, $1 \, \sigma$ range and pull with respect to the SM values are shown in the second, third and fourth
          column, respectively.}
            {\begin{tabular}{@{}cccc@{}} \toprule
                {\rm Coefficient} & {\rm Best Fit} &  $1\, \sigma$  & {\rm Pull}$_{\rm SM}$ \\
                \hline
                $C^{\rm NP}_7$        & -0.02    & [-0.04,0.00]    & 1.2 \\
                \boldmath{$ C^{\rm NP}_9$} & { -1.09}  & { [-1.29,-0.87]} & {\bf 4.5} \\
                $C^{\rm NP}_{10}$      & 0.56   & [0.32,0.81]   & 2.5 \\
                $C^{'\rm NP}_{7}$      & 0.02   & [-0.01,0.04]  & 0.6 \\
                $C^{'\rm NP}_{9}$      & 0.46   & [0.18,0.74]   & 1.7 \\
                $C^{'\rm NP}_{10}$     & -0.25  & [-0.44,-0.06]   & 1.3 \\
                $C^{\rm NP}_{9} = C^{\rm NP}_{10}$ & -0.22 & [-0.40,-0.02] & 1.1 \\
                \boldmath{$C^{\rm NP}_{9} = - C^{\rm NP}_{10}$} & {\bf -0.68} & { [-0.85,-0.50]} & {\bf 4.2} \\
                \boldmath{$C^{\rm NP}_{9} = - C^{'\rm NP}_{9}$} & {\bf -1.06} & {[-0.85,-0.50]} & {\bf 4.8} \\
                \hline
              \end{tabular} \label{tab:C}}
      \end{table}
      
      The agreement of the results of the fit with respect to different theory inputs
      (as for example, Ref.~\refcite{theory_AFB} and Ref.~\refcite{Descotes_04239})
      show that they are robust against the chosen methodology.
      Unfortunately the fact that most of the NP contribution can be accommodated in a non-zero value for
      $C^{\rm NP}_{9}$, prevents from an unambiguous interpretation of the results, as non perturbative
      charm-loop effects could mimic the same effect.

\begin{figure}[h]
        \centerline{\includegraphics[width=0.65\linewidth]{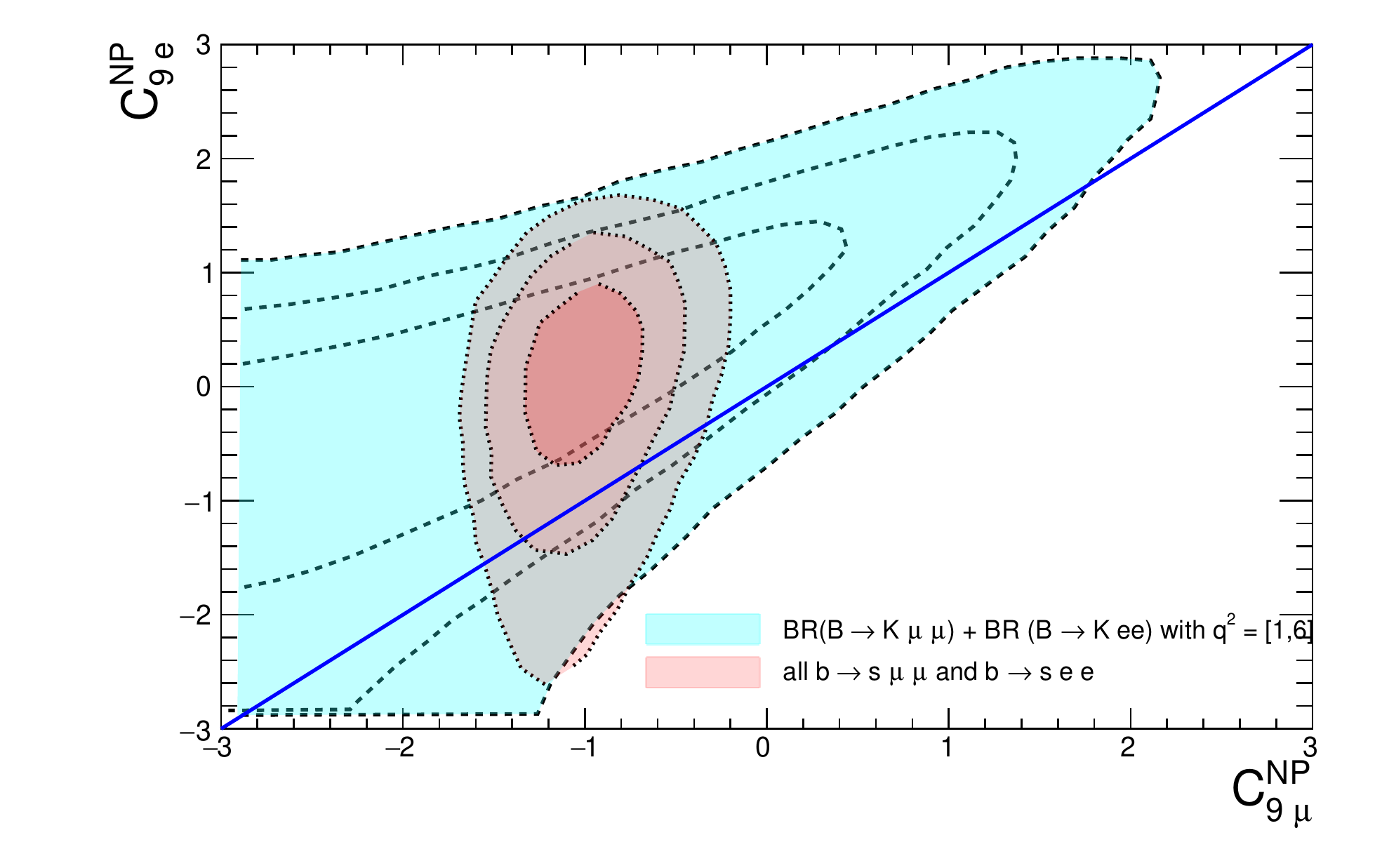}}
        \centerline{\includegraphics[width=0.65\linewidth]{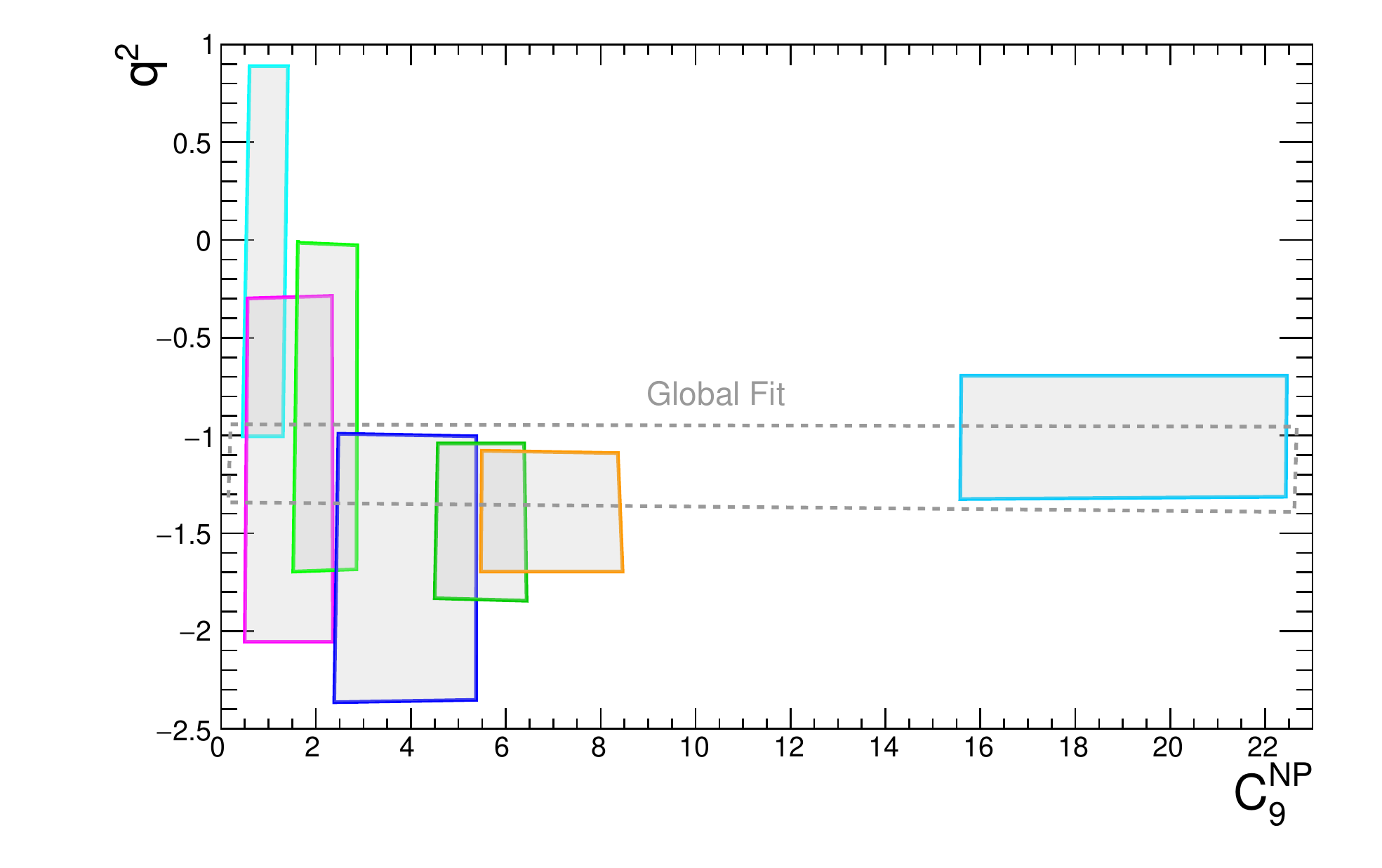}}
        \vspace*{8pt}
        \caption{Upper plot: fit which allows for LFU violating contributions in independent coefficients $C^{\rm NP}_{9 \mu}$ and
          $C^{\rm NP}_{9 e}$. Bottom plot: fit of a single parameter scenario in different bins of $q^2$
          with a single coefficient $C^{\rm NP}_{9}$.
          Reproduced from Ref.~\cite{Descotes_04239}.
          \protect\label{fig:C9_q2}}
      \end{figure}

      \vskip 2mm
      A promising method to resolve the origin of the deviation of $C_9$ from SM predictions is the analysis
      of its $q^2$ dependence. In fact, a high-energy NP contribution would hardly generate any $q^2$
      dependence of $C^{(')}_9$ while a $q^2$ dependence is expected if this deviation is caused by charm-loops effects.
      The analysis of a possible $q^2$ dependence of $C_9^{\rm NP}$ has been performed with different
      methods~\cite{Altmannshofer_06199,Descotes_04239} and the results are in agreement with each other,
      as shown in Fig.~\ref{fig:C9_q2}.

      Unfortunately the situation is still not clear due to the large statistical uncertainties of each bin that
      do not allow to exclude any of the two hypotheses.
      It is worth noting that long-distance charm-loop effects in $b \to s \ell^+ \ell^-$
      transitions are lepton-flavor conserving processes and therefore cannot be the origin of
      LFU violation effects observed in the measurement
      of $R_K$ and $R_{K^*}$.

      The experimental results on $R_K$ and $R_{K^*}$ suggest a possible violation of the LFU, hence a different NP
      contributions for $C^{(')}_{9\mu}$ and $C^{(')}_{9 e}$.  Figure~\ref{fig:C9_q2}
      shows the result of the fit assuming two separate
      contribution to $C^{(')}_{9\mu}$ and $C^{(')}_{9e}$.
      The data suggest NP contribution only associated to the coupling with muons.
      However it is puzzling how NP could contribute to the deviation of $R_{K^{0*}}$ in the bin $q^2=[1.1,6]$ GeV$^{2}$/c$^4$.
      In fact, at low $q^2$ SM contribution is dominated by the dipole operator associated to the photon pole and NP effects
      are expected to be highly reduced in this bin, unless light long-range NP effects are present~\cite{Altmannshofer_07494}.

      \vskip 2mm 
      Different attempts have been tried to explain this coherent set of anomalies in terms of NP
      affecting the $O_9$ and $O_{10}$ operators.
      NP can contribute at tree level, via the exchange of a heavy vector-boson
      $Z^{'}$~\cite{Descotes_5683, Buras_2466, Gauld_1082, Altmannshofer_1269, Celis_03079, Falkowski_01249,
       Crivellin_00993,Crivellin_03477, Crivellin_07928, Beciveric_00881, Crivellin_02703, Faroughy_07138},
       or of a scalar or vector
       lepto-quark~\cite{Hiller_1627, Gripaios_1791, Beciveric_09024, DeMedeiros_01084, Fajfer, Beciveric_08501, DiLuzio_08450},
       or at the loop level via box diagrams including new particles~\cite{Gripaios_05020, Bauer_01900, Arnan_07832}
      or $Z^{'}$ penguin diagrams~\cite{Belanger_06660}.
      
      A popular category of $Z'$ models relies on gauging $L_{\tau}-L_{\mu}$ lepton
      number~\cite{Altmannshofer_1269,Crivellin_00993, Crivellin_03477, Altmannshofer_07009, Altmannshofer_08221}.
      These models are well suited to explain LFU violations in $R_K$ and $R_K^*$ given the vanishing coupling to
      electrons of the $Z'$.
      Moreover, the inclusion of the $Z'$ into a $SU(2)'$ gauge symmetry, allows also to explain
      the $R_D$ and $R_D^*$ anomalies, via a tree-level exchange of a $W'-$boson in
      the $b \to c \ell^- \overline{\nu}$ transitions~\cite{Boucenna_01349,Greljo_01705}
      If the $Z'$ coupling to muons occurs at the loop level,
      NP contributions responsible of the $b \to c \ell^+ \ell^-$ anomalies could possibly also explain the $(g-2)_{\mu}$
      anomaly~\cite{Belanger_06660}.
      Models with lepto-quarks that couple leptons to quarks at the tree level
      intrinsically violate LFU and are therefore excellent candidates to accommodate LFU violating measurements
      as $R_K$, $R_{K^*}$ and $R_D^{(*)}$~\cite{Hiller_1627,Gripaios_1791, Beciveric_09024, DeMedeiros_01084, Fajfer, Beciveric_08501}.
      Moreover lepto-quark models induce tree-level transition
      in $b \to s \ell^+  \ell^- $ processes and would therefore escape constraints from $B^0_s -\overline{B^0}_s$
      mixing, where they would contribute only at the loop level.
      On the other hand direct
      searches~\cite{cms_1703.03995,cms_01190, cms_PAS-EXO-16-043, cms_PAS-EXO-17-003, cms_PAS-SUS-18-001, cms_PAS-EXO-17-029, cms_02864}
      already severely constrain some of
      these models and more are expected to be challenged in the near future.

      \section{Search for light and weakly coupled particles in rare b decays}
      
      The lack of any evidence for new particles in
      LHC collisions has rekindled the interest in hidden sector theories (see for example Ref.~\refcite{ship_PP})
      In contrast to
      most SM extensions, these theories postulate that dark matter particles do not carry SM charges and couple
      only feebly to the SM particles. Hidden sector particles are singlet states under the SM gauge interactions and
      couple very feebly to the SM particles.
      The couplings arise via mixing of the hidden-sector field with a
      SM ``portal'' operator.

      In general, the renormalisable portals with lower dimension in the SM can be classified as follows:
      
      \begin{center}
        \begin{tabular}{rl}
          Portal & Coupling \\
          \hline
          \smallskip
          Dark Photon, $A_{\mu}$ &  $-\tfrac{\epsilon}{2 \cos\theta_W} F'_{\mu\nu} B^{\mu\nu}$ \\ \smallskip
          Dark Higgs, $S$  & $(\mu S + \lambda S^2) H^{\dagger} H$  \\ \smallskip
          Axion, $a$ &  $\tfrac{a}{f_a} F_{\mu\nu}  \tilde{F}^{\mu\nu}$, $\tfrac{a}{f_a} G_{i, \mu\nu}  \tilde{G}^{\mu\nu}_{i}$ \\ \smallskip
          Sterile Neutrino, $N$ & $y_N L H N$ \\
        \end{tabular}
      \end{center}
      
      \noindent
      where $F'_{\mu \nu}$ is the field strength for the dark photon, which couples to the hypercharge field,
      $B^{\mu\nu}$; $S$ is a new scalar singlet that couples to the Higgs doublet, $H$, with dimensionless and
      dimensional couplings, $\lambda$ and $\mu$; $a$ is a pseudoscalar axion that couples to a dimension-4 diphoton
      or digluon operator; and $N$ is a new neutral fermion that couples to one of the left-handed doublets
      of the SM and the Higgs field with a Yukawa coupling $y_N$.

      A lively and broad experimental activity to search for hidden sector particles is currently taking place
      worldwide (for a review see Ref.~\refcite{US_cosmic_vision}).
      Several experimental technique are considered depending on the mass range of the
      hidden particles and related mediators.
      Decays of $b$ hadrons are a unique probe for dark-sector models with particles at the GeV mass scale.
      In particular FCNC transitions as  $b \to  s \ell^+ \ell^-$ can be
      sensitive to light scalar ($\chi$) or pseudo-scalar ($a$)  mediators
      via the process $b \to s \chi$, with $\chi \to \ell^+ \ell^-$.
      
      The LHCb experiment has published a search for a light scalar or pseudo-scalar particle $\chi$ produced in the decays
      $B^0 \to K^{*0} \chi$, $K^{0*} \to K^+ \pi^-$ and $\chi \to \mu^+ \mu^-$ and $B^+ \to K^+ \chi$, $\chi \to \mu^+ \mu^-$,
      by looking at peaks in the dimuon invariant mass.
      The analyses were both based on the full Run 1 dataset, 3 fb$^{-1}$ ~\cite{lhcb_darkboson1, lhcb_darkboson2}.
      In both cases no signal has been observed and upper limits on the branching fraction have been set as a function of the mass and the lifetime
      of the light hidden mediator. 
      These limits are of the order of $10^{-9}$ for lifetimes less than 100 ps and
      for $m_{\mu \mu} < 4.5$ GeV.

      \section{Conclusions}

      With the discovery of the Higgs boson with a mass of 125 GeV and with the absence of unambiguous
      signal of NP in direct searches at the LHC, our concept of naturalness is currently under pressure.
      Perhaps NP is at a mass scale much higher than that directly accessible at present colliders
      and can be detected only via measuring deviations in precision measurements with respect to SM predictions,
      or it is below the EW scale, couples very weakly to the SM world and so far escaped detection.
      FCNC b-hadron decays are an excellent probe for exploring both directions.
      A wealth of experimental measurements and theoretical computations
      in the last decade allowed the flavor community to test the CKM paradigm with unprecedented precision.

      \vskip 2mm
      Everywhere an excellent agreement with the SM expectations has been found except for a set of intriguing anomalies
      observed in the measurements of branching fractions, ratios of branching fractions and angular distributions
      in FCNC $b \to s \ell^+ \ell^-$ transitions and ratios of branching fractions in $b \to c \tau/\mu \overline{\nu}$ transitions.
      
      Several correlated analyses of these anomalies have been performed in order to identify a
      possible universal NP contribution that could provide a coherent description of the data and this activity 
      will surely continue in the future.

      \vskip 2mm
      From the experimental point of view, the data already collected by the LHC experiments during Run 2 will allow to reduce
      significantly the uncertainty of all the measurements contributing ot the anomalies, that are currently statistically limited,
      and increase (or decrease) the tension with the SM predictions.
      Moreover Belle II is expected to start data taking with its full detector in 2018 and aims to collect an
      integrated luminosity of 50 ab$^{-1}$ by 2024. This will provide a dataset
      that is about a factor of 50 times larger than that collected by BaBar and Belle together and will open the possibility to confirm
      (or disprove) the tensions observed in $b \to s \ell^+ \ell^-$ and $b \to c \ell \overline{\nu}$ transitions.

      In the coming decade, FCNC $b$-hadrons decays will continue to play a central role in testing the SM picture and 
      in setting up new directions for model building for NP contributions, in particular if no sign of NP will be found
      in direct searches at the LHC.

\section*{Acknowledgments}
The author warmly thank the members of the LHCb collaboration and
in particular  Marc-Olivier Bettler for his careful reading of the manuscript and interesting comments and discussions.



\end{document}